\documentclass[fleqn,usenatbib]{mnras}

\usepackage{newtxtext,newtxmath}

\newcommand{\mpch}{h^{-1}\>{\rm{Mpc}}}

\newcommand{\Omegam}{\>{\Omega_\mathrm{m}}}
\newcommand{\fb}{\>{f_\mathrm{b}}}
\newcommand{\h}{\>h}
\newcommand{\ns}{\>n_\mathrm{s}} 
\newcommand{\As}{\>A_\mathrm{s}}
\newcommand{\wa}{\>{w_a}}
\newcommand{\SigmaMnu}{\>{M_{\nu}}}
\newcommand{\alphas}{\>\rm{\alpha_s}}
\newcommand{\Gammadcdm}{\rm{\Gamma_{dcdm}}}
\newcommand{\mpc}{\>{\rm {Mpc}}}

\usepackage[T1]{fontenc}

\DeclareRobustCommand{\VAN}[3]{#2}
\let\VANthebibliography\thebibliography
\def\thebibliography{\DeclareRobustCommand{\VAN}[3]{##3}\VANthebibliography}

\usepackage{graphicx}	
\usepackage{amsmath}	

\usepackage{subcaption}
\usepackage{xcolor}

\graphicspath{{Figures/}}

\title[Neutrino mass from persistent homology]{Revealing the neutrino mass through persistent homology of the cosmic~web}

\author[J. Wang et al.]{Jiaqi Wang,$^{1,2}$\thanks{E-mail: jiaqi.wang@durham.ac.uk, wangjia777@sjtu.edu.cn}
Willem Elbers,$^{2}$\thanks{E-mail: willem.h.elbers@durham.ac.uk}
Carlos S. Frenk,$^{2}$
Shaun Cole,$^{2}$
Xiaohu Yang,$^{1,3}$
Ian G. McCarthy,$^{4}$\newauthor
Rien van de Weygaert$^{5}$
\\
$^{1}$State Key Laboratory of Dark Matter Physics, Tsung-Dao Lee Institute \& School of Physics and Astronomy, Shanghai Jiao Tong University, Shanghai \\ 201210, China\\
$^{2}$Institute for Computational Cosmology, Department of Physics, Durham University, South Road, Durham, DH1 3LE, UK\\
$^{3}$Shanghai Key Laboratory for Particle Physics and Cosmology, and Key Laboratory for Particle Physics, Astrophysics and Cosmology, Ministry of \\Education, Shanghai Jiao Tong University, Shanghai 200240, China\\
$^{4}$Astrophysics Research Institute, Liverpool John Moores University, 146 Brownlow Hill, Liverpool L3 5RF, UK\\
$^{5}$Kapteyn Astronomical Institute, University of Groningen, PO Box 800, 9700 AV Groningen, The Netherlands
}

\date{Accepted XXX. Received YYY; in original form ZZZ}

\pubyear{2021}
\begin{document}
\label{firstpage}
\pagerange{\pageref{firstpage}-\pageref{lastpage}}
\maketitle

\begin{abstract}
Cosmological constraints on neutrino mass offer a promising avenue for advancing our understanding of both fundamental particle physics and the evolution of cosmic large-scale structure.
To overcome challenges associated with traditional probes of neutrino mass, particularly degeneracies with other parameters, we consider topological summaries of the cosmic web based on the formalism of persistent homology. We introduce persistence strips, a novel representation that segments Betti curves by topological persistence, effectively condensing two‑dimensional persistence diagrams into a set of one‑dimensional curves.
Applied to the FLAMINGO suite of cosmological simulations, these topological descriptors demonstrate pronounced sensitivity to neutrino mass.
By constructing an emulator spanning a 10-dimensional $w_0 w_a\text{CDM} +\nu$ cosmological parameter space that includes parameters degenerate with neutrino masses in conventional approaches,
assuming a volume of only $(350 \mpc)^3$,
we obtain neutrino mass constraints with an uncertainty of 0.05 eV for the total matter field and 0.13 eV for the dark matter-only field, with the strongest constraints coming from void topology.
Persistence strips exhibit roughly twice the constraining power of unbinned Betti curves and, through their multi-scale, environment-dependent sensitivity, systematically break degeneracies between neutrino mass and other cosmological parameters. 
We pinpoint the precise physical origin of the signal, identifying two equally important contributions: sensitivity to the neutrino mass fraction, which is highest in underdense regions, and the impact of neutrinos on the distribution of dark matter.
Our findings highlight the particular promise of applying topological statistics to weak lensing measurements, which directly probe the total matter distribution. 
\end{abstract}

\begin{keywords}
cosmology: theory -- dark matter -- large-scale structure of universe, physical data and processes: neutrinos
\end{keywords}



\section{Introduction}
\label{sec:intro}
Massive neutrinos represent a crucial extension of the standard model of particle physics and play a vital role in the evolution of the Universe.
As the only known form of hot dark matter, they exhibit key differences from the cold dark matter (CDM) that makes up most of the matter budget in the standard cosmological paradigm.
While CDM clumps under gravity to form cosmic structures, neutrinos suppress the growth of structure on small scales. Their significant thermal motions prevent them from clustering effectively, thereby smoothing out the matter distribution \citep{Bond80,Hu_Neugalaxy_1998,Brandbyge_Neu_thermalmotion_2008}. In addition, neutrinos alter the cosmic expansion history by behaving as non-relativistic matter only at late times. Consequently, cosmological observations provide powerful and complementary constraints on the total neutrino mass, $\sum m_\nu$, to those derived from laboratory experiments \citep[e.g.][]{Abazajian_Neu_astro_2011,Esteban24_NuFit}.

The cosmological imprints of neutrino mass can be probed through a variety of astrophysical observables, usually by combining probes of the cosmic large-scale structure with information derived from the cosmic microwave background (CMB), where the latter provides an anchor at large scales and early times when neutrinos are still relativistic \citep[e.g.][]{Abazajian_Neu_CMB_2015}. The strongest current constraints are derived from CMB data from the \emph{Planck} satellite \citep{Planck2018} and the ground-based Atacama Cosmology and South Pole Telescopes \citep{ACT25_DR6,SPT25_3GD1}, combined with measurements of baryon acoustic oscillations (BAO) in the distribution of galaxies, quasars, and the Lyman-$\alpha$ forest from the second data release of the Dark Energy Spectroscopic Instrument \citep{DESI-DR2-Lya2025,DESI-DR2cosmology2025}. When analyzed within the standard $\Lambda$CDM framework, these data provide $95\%$ upper bounds in the range $\sum m_\nu<0.04-0.06\,\mathrm{eV}$ \citep{Elbers25_DESI,SPT25_3GD1}. This is close to or below the lower bounds derived from terrestrial neutrino oscillation measurements, $\sum m_\nu>0.059\,\mathrm{eV}$, giving rise to a neutrino mass tension that could point to new neutrino physics or a non-standard cosmological evolution \citep[e.g.][]{Craig24,elbers_negative_2024,Elbers25_DESI,Green_Neg_Neu_2025,NaredoTuero24,Noriega24,Avsajanishvili_Neuvarying_2025,Jiang25}. While conventional CMB and BAO techniques currently provide the strongest and most reliable constraints on $\sum m_\nu$, the latest hints of a neutrino mass tension motivate the development of alternative approaches that could add complementary information.

Additional information can be extracted by modelling the shape of the power spectrum of density fluctuations probed by galaxy clustering \citep[e.g.][]{Hu_Neugalaxy_1998,Elgaroy02,Tegmark04}, weak lensing \citep[e.g.][]{Cooray_Neu_WL_1999, Abazajian_Neu_WL_2003}, and the absorption spectra of the Lyman-$\alpha$ forest \citep[e.g.][]{Croft99,Seljak05}.
While these probes provide valuable information, their constraining power is subject to challenges, including theoretical modelling uncertainties, degeneracies with other parameters (particularly the amplitude of matter fluctuations, $\sigma_8$~\footnote{Defined as the standard deviation of the present-day linear matter field averaged in spheres of radius $8h^{-1}\,\rm{Mpc}$.}, and the spectral index, $n_\mathrm{s}$), and limited signal-to-noise ratios in current observations.

Recognizing these limitations, several studies have proposed that higher-order statistics could provide improved constraints on neutrino masses \citep{Chiang_Neu_bispectrum_2018, Hahn2021}. For instance, \citet{Hahn2021} demonstrated that the bispectrum can supplement information from the power spectrum and help break the degeneracy between $\sum m_\nu$ and $\sigma_8$, despite the challenges associated with its modelling. However, the utility of such statistics may be limited in certain contexts; \citet{Bayer_Fake_Neu_2022} argued that for 2D probes like the halo field and weak lensing at scales $k \leq 1 h\,\mathrm{Mpc}^{-1}$, higher-order statistics contribute little information beyond the linear power spectrum. Although the 3D matter field may contain more information, obtaining high-quality observational data remains difficult.

An alternative strategy involves focusing on specific cosmic environments, such as voids \citep[e.g.][]{Massara15, Kreisch_Neufingerprint_2019, Thiele_Neu_void_2024}.
Nevertheless, these methods face challenges: results are sensitive to void definitions that vary across studies, and analytical models of void statistics often rely on simplifying assumptions that limit their applicability to the full void population. These issues underscore the need for an effective and unambiguous summary statistic that balances conciseness with informational richness. For example, \citet{Bayer2024} showed that the halo mass function in environments split according to the emptiness can yield tight bounds on neutrino masses.

Recent years have seen the emergence of several novel cosmological probes that target the geometry or topology of the cosmic web, including the minimum spanning tree \citep{Naidoo_miniSTree_2022,Simoes25_MST}, $k$-Nearest Neighbour distributions ($k$NN) \citep{Gangopadhyay_kNN_connection_2025, Ouellette_betti_kNN_2025}, the clustering of cosmic web components \citep{Sunseri25}, and topological data analysis \citep[TDA hereafter]{van_de_weygaert_alpha_2011}. Among these, TDA has shown particular promise as a powerful tool for characterizing the multiscale topological structure of the cosmic web. Through persistent homology, a method that quantifies the evolution of topological features across a range of physical scales or density thresholds, TDA provides a rigorous framework for describing the ``shape" of large-scale structure \citep{Carlsson_Topology_2009}. This approach encodes information in the form of persistence diagrams, which track the birth and death scales of topological features and contain significantly more information than the traditional Euler characteristic. Moreover, persistent homology captures structural aspects of the density field complementary to conventional $N$-point statistics and offers a natural definition of the signal-to-noise ratio through the scale-persistence relationship of features in the persistence diagrams (PDs hereafter) \citep{Pranav_topologyweb_2017}. TDA has found diverse applications in astrophysics, including the detection of non-Gaussianity \citep{Biagetti_nonGaussianity_2021}, the analysis of the topology of reionization \citep{elbers_persistent_2019, Elbers_reionizationtopo_2023, Giri21}, constraints on redshift-space distortions \citep{Liu_DTFE_RSD_2024}, and the extraction of cosmological information from cosmic shear \citep[e.g.][]{Heydenreich_cosmicshear_2021}. Within large-scale structure studies, TDA offers a powerful framework for quantifying the complex topology of the cosmic web \citep{van_de_weygaert_alpha_2011} and has been used to constrain both cosmological parameters and galaxy evolution processes \citep{Yip_cosmotopology_2024}. Common topological methods include the Euler characteristic \citep{van_de_weygaert_alpha_2011}, the $\beta$-skeleton \citep{Fang_betaskeleton_2019}, Minkowski functionals \citep{Liu_NeuMFS_2020, Liu_MFsneu_2022}, critical points \citep{moon_signature_2023}, and persistent homology \citep{Wilding_topoweb_2021, Tsizh_Wassdist_2023}. The last approach characterizes multiscale topology through Betti curves, which track the evolution of topological features, such as connected components, loops, and voids, across a hierarchy of filtration parameter.

Recent work by \cite{jalalikanafi_imprint_2024} has demonstrated the particular value of Betti curves in constraining neutrino mass. Through Fisher forecasts using simulations with varying $\sum m_\nu$ and $\sigma_8$ (with other parameters held fixed), they showed that Betti curves not only provide tight constraints on $\sum m_\nu$ but also help break its degeneracy with $\sigma_8$. Notably, they found that the degeneracy-breaking capability differs between the total matter field (including neutrinos) and the combined field of dark matter and baryons---a phenomenon whose physical origin warrants further investigation.
To extend this research, we consider a broader cosmological parameter space, including $\Omega_\mathrm{m}$ and $h$, and utilize higher-resolution neutrino simulations from the FLAMINGO suite \citep{Kugel23,schaye_flamingo_2023}, which enable exploration of smaller scales than previously possible with Quijote \citep{Quijote_Navarro_2020}. This approach allows us to derive more comprehensive parameter constraints and better understand the scale-dependent information content of topological summaries.

To address these challenges, we conduct a comprehensive investigation across a high-dimensional cosmological parameter space. However, modelling topological statistics presents significant difficulties, particularly in such an expansive parameter regime where analytical approaches become intractable. To overcome this limitation, we employ emulation techniques, constructing surrogate models that accurately replicate the output of full-scale cosmological simulations at a fraction of the computational cost. This approach offers dual advantages in model fitting: dramatically reduced computation times and enabling the use of sophisticated statistical methods including gradient-based optimization and efficient sampling algorithms that would otherwise be prohibitively expensive. Emulation has demonstrated remarkable success across cosmological applications, providing an effective methodology for balancing precision and computational efficiency in data-intensive studies.

In this work, we aim to characterize the multiscale topological structure and connectivity of the cosmic web using persistent homology and Betti curves. We develop an emulator capable of predicting topological statistics across a 10-dimensional cosmological parameter space and check the cosmological parameter inference. 
The paper is organized as follows: Section~\ref{sec:data} describes the simulations used in our analysis. Section~\ref{sec:homo} introduces the theoretical framework of TDA and examines the response of Betti curves to neutrino mass variations. 
Section~\ref{sec:emu} details the emulator design and training. 
Section~\ref{sec:perfm} presents our main results on parameter dependence and cosmological constraints. 
Section~\ref{sec:discuss} discusses the physical origin of the signals.
Section~\ref{sec:obs} checks some of the observational effects and presents further discussion.
Finally, Section~\ref{sec:summary} summarizes our findings and conclusions. 
Note that in this paper $\log$ is in base $10$. 

\section{Data} \label{sec:data}

We make use of two sets of simulations in our analysis: a subset of simulations from the FLAMINGO suite (Section~\ref{sec:FLAMINGO}) and a larger set of 110 simulations spanning a cosmological parameter grid (Section~\ref{sec:data_emu}). The first are used for a systematic exploration of the topological effects of massive neutrinos, while the second are used to construct our persistent homology emulator.

\subsection{FLAMINGO simulations}\label{sec:FLAMINGO}
The Virgo Consortium’s FLAMINGO suite represents a state-of-the-art set of cosmological simulations designed to advance our understanding of structure formation, galaxy cluster physics, and the role of neutrinos and baryons in cosmic evolution \citep[][Helly et al. in prep.]{Kugel23,schaye_flamingo_2023}. The original suite contains 12 dark matter only simulations and 16 hydrodynamical simulations, which were later complemented with additional neutrino mass variations and extensions with decaying dark matter \citep{elbers_flamingo_2024}. The suite comprises large simulations in multiple volumes and resolutions, incorporating both gravitational physics and detailed prescriptions for galaxy formation. While FLAMINGO offers many baryonic physics variations, we do not use the hydrodynamical simulations in this study, where our focus is on the gravitational impact of neutrinos. We exploit the detailed implementation of massive neutrinos in the FLAMINGO simulations to investigate the constraining power of TDA on neutrino mass, as well as its degeneracies with other cosmological parameters.

The FLAMINGO simulations were run with the modern and highly scalable cosmological hydrodynamics code SWIFT \citep{Schaller18,Schaller23}. Massive neutrinos are included using the computationally efficient $\delta f$ method \citep{elbers_optimal_2021}, which significantly reduces shot noise and allows for accurate, self-consistent evolution of neutrino perturbations. The simulations are started from higher-order Lagrangian multi-fluid initial conditions, incorporating corrections for the presence of massive neutrinos \citep{Hahn21,Elbers22,Elbers22b}.

For this study, we primarily rely on FLAMINGO simulations with a $1\,\mathrm{Gpc}^3$ volume, consisting of $1800^3$ cold dark matter particles and $1000^3$ neutrino particles, for which multiple neutrino mass variations are available. We also use the larger $2.8\,\mathrm{Gpc}^3$ simulation with $5040^3$ cold dark matter particles and $2800^3$ neutrino particles to estimate the covariance matrix of the statistical errors of our estimates. To explore the effect of neutrino mass, we use three different values: $0.06~\text{eV}$ (the baseline), $0.24~\text{eV}$, and $0.48~\text{eV}$. This range enables a detailed examination of the imprint of neutrino mass on the topological structure of the cosmic web, including its interplay with other cosmological parameters.

\subsection{The emulator}\label{sec:data_emu}

In order to constrain cosmological parameters including the total mass of neutrinos, we developed an emulator to model our topological measurements.
The emulator is based on a suite of 50 cosmological $N$-body simulations with a side length of $L = 350\,\mathrm{Mpc}$ and 50 additional simulations with a side length of $L = 700\,\mathrm{Mpc}$, each containing $1260^3$ cold dark matter particles and $700^3$ neutrino particles, evolved under periodic boundary conditions. The simulations were run with the SWIFT code and their treatment of neutrinos was identical to the FLAMINGO simulations described above.

The simulations are designed for the sampling of 10 cosmological parameters: \{$\Omega_\textrm{m}$, $\fb$, $\h$, $\ns$, $\As$, $w_0$, $\wa$, $\SigmaMnu$, $\alphas$, $\Gammadcdm$\}, where the parameter ranges are shown in Table~\ref{tab:para-Latin}. This includes the standard $\Lambda$CDM parameters: the matter density $\Omega_\mathrm{m}$, baryon fraction $f_\mathrm{b}\equiv\Omega_\mathrm{b}/\Omega_\mathrm{m}$, Hubble parameter $h$, power spectrum amplitude $A_\mathrm{s}$, and spectral index $n_\mathrm{s}$. Beyond these parameters, the emulator incorporates extensions for dynamical dark energy with an evolving equation of state parametrized by $w_0$ and $w_a$ \citep{Chevallier2001,Linder2003}, non-zero neutrino masses $M_\nu\equiv\sum m_\nu$, and running of the spectral index $\alpha_\mathrm{s}$. 
Finally, $\Gamma_\textrm{dcdm}$ indicates the decay rate of cold dark matter in units of $\mathrm{km}\,\mathrm{s}^{-1}\,\mathrm{Mpc}^{-1}$, as described in \citet{elbers_flamingo_2024}. The parameters were sampled according to a space-filling orthogonal Latin hypercube, but points were transformed in the $(w_0,w_a)$-plane to primarily sample the space around the degeneracy direction inferred from the DESI DR1 results \citep{DESI_DR1_Cosmology}. 
To evaluate the performance in the presence of neutrino mass, four additional simulations with $L = 350\,\mathrm{Mpc}$ are run with varying neutrino mass while holding other parameters fixed.
This simulation suite will be described in greater detail in forthcoming publications (McCarthy et al., in prep.).

\begin{table}
    \centering
    \caption{Allowed ranges of each cosmological parameter in our emulator. The upper rows show the 10 original parameters used to construct the hypercube. We also list the range of the $\sigma_8$ parameter used in this study.
    }
    \begin{tabular}{ccc}
    \hline
      parameter& minimum & maximum \\
      \hline
        $\Omegam$ & 0.17 & 0.38\\
        $f_\mathrm{b}$ & 0.14 & 0.17 \\
        $h$ & 0.6 & 0.8\\
        $n_\mathrm{s}$& 0.9 & 1\\
        $\ln (10^{10}A_\mathrm{s})$ & 2.97 & 3.11\\
        $w_0$ & -1.19 & -0.30\\
        $w_a$ & -2.74 & 1.08 \\
        $\SigmaMnu [\rm eV]$ & 0.005 & 0.45\\
        $\alphas$ & -0.03 & 0.03\\
        $\Gamma_\textrm{dcdm} \mathrm{[10^{-3} Gyr^{-1}]}$ & 0.33 & 19.8\\
        \hline
        $\rm \sigma_8$ & 0.48 & 1.05\\
        \hline
    \end{tabular}
    \label{tab:para-Latin}
\end{table}
\subsection{Density contrast}

In order to save computational costs, we focus in this study on the density field as input data and employ the density contrast $1+\delta(\boldsymbol{r}) \equiv {n(\boldsymbol{r})}/{\bar{n}}$ as our filtration (in Section~\ref{persistenthomology}).
The density contrast for the total matter field is defined as,
\begin{equation}
\delta_\mathrm{m} \equiv \frac{\Omega_{\mathrm{CDM}}}{\Omega_{\mathrm{m}}} \delta_{\mathrm{CDM}}
+\frac{\Omega_{\mathrm{b}}}{\Omega_{\mathrm{m}}} \delta_{\mathrm{b}}
+\frac{\Omega_\nu}{\Omega_{\mathrm{m}}} \delta_\nu = \frac{\Omega_{\mathrm{CDM}}+\Omega_{\mathrm{b}}}{\Omega_{\mathrm{m}}} \delta_{\mathrm{CDM}}
+\frac{\Omega_\nu}{\Omega_{\mathrm{m}}} \delta_\nu  ,
\end{equation}
where $\Omega_{\mathrm{m}} = \Omega_{\mathrm{CDM}} +\Omega_{\mathrm{b}}+\Omega_{\mathrm{\nu}}$. The second equality comes from the fact that we assume the baryonic field to be the same as the CDM field since we do not use hydrodynamical simulations.

We build the density contrast fields from simulation snapshots for different mass components using the Pylians \citep{Pylians} package\footnote{\url{https://pylians3.readthedocs.io/en/master/index.html}}
on a $512^3$ grid using the Cloud-In-Cell (CIC hereafter) mass-assignment scheme \citep{Laux_CIC_1996}.
We generally consider smoothed fields in our analysis, convolving the underlying field with a Gaussian window function.
As described in Section~\ref{sec:smoothing}, we consider a range of different smoothing scales, with a fiducial choice of $ R = 2\, \mpc$.

\begin{figure*}
    \centering
    \includegraphics[width=0.95\linewidth]{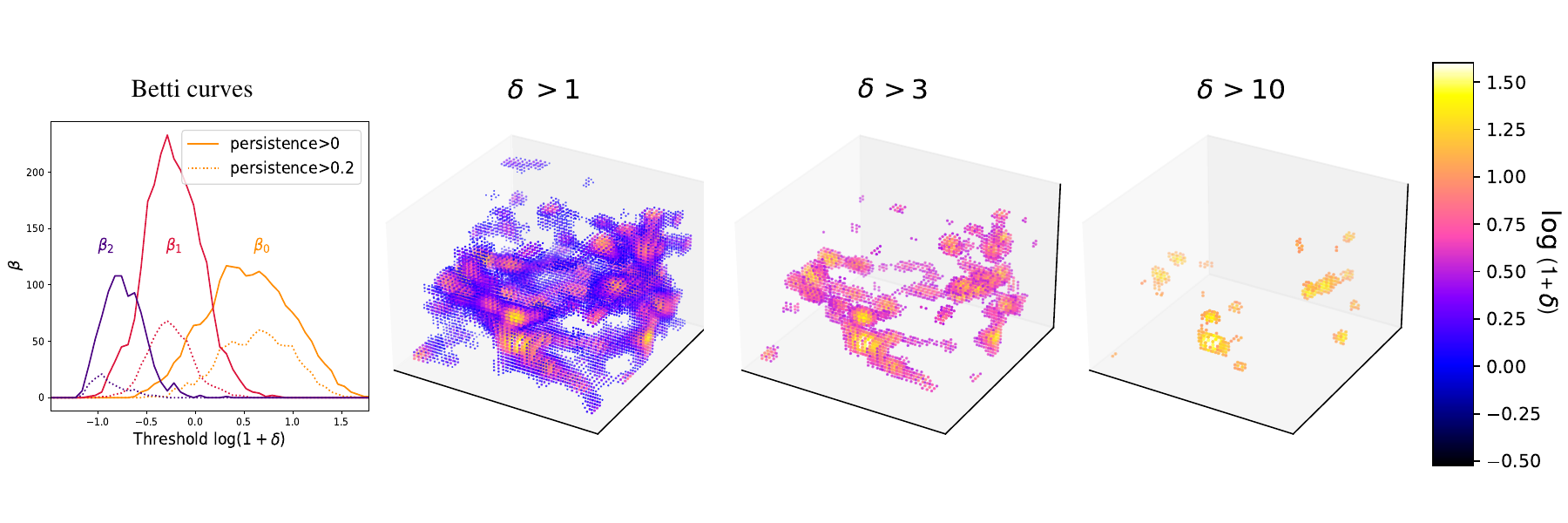}
    \caption{Topological structure of a 3D cosmic density field. (Left) Betti curves $\beta_k$ as a function of density threshold, showing the number of $k$-dimensional topological features. From left to right: cosmic voids ($\beta_2$, purple), filamentary tunnels ($\beta_1$, red), and halos ($\beta_0$, orange). Solid and dotted lines correspond to persistence thresholds of 0 and 0.2, respectively, applied to the underlying persistence diagrams. (Right) Visualizations of the identified structures at different density thresholds. Colour in the 3D renderings represents the local matter overdensity, as indicated by the colour bar.
    }
    \label{fig:dens_example}
\end{figure*}

\section{Persistent homology and cosmic web}
\label{sec:homo} 

The conceptual foundation for understanding walls, filaments, and voids in the large-scale structure was established by the Zeldovich school through the Zeldovich approximation \citep{Zeldovich70} and the adhesion model \citep{ShandarinZeldovich89}. The term `cosmic web' was later coined by \citet{Bond96} to emphasize the interconnected, hierarchical nature of this structure. Subsequent theoretical work has further elucidated the multiscale evolution of voids \citep{Sahni1994, Sheth2004}, the broader cosmic web \citep{Shen2006}, filaments \citep{Feldbrugge2025}, and walls \citep{Hertzsch2026}.

Topology provides a complementary perspective by characterizing this structure as a manifold, focusing on its intrinsic connectivity rather than its precise metric properties~\citep{Edelsbrunner2009ComputationalTA, Wasserman2018, Carlsson_Vejdemo-Johansson_2021}. Persistent homology enables the systematic tracking of the emergence and disappearance of structures classified as $k$-cycles: connected components (0-cycles), loops enclosing tunnels (1-cycles), and shells enclosing cavities (2-cycles), which correspond to halos, filaments, and voids, respectively (see \citealt{Munkres84,Hatcher01}). 
Throughout this study, we calculate the topological properties of the field data using the python package GUDHI \citep{gudhi:urm, gudhi:CubicalComplex}\footnote{\url{https://gudhi.inria.fr/python/latest/}}. The field data are represented as a cubical complex with periodic boundary conditions, from which the topological properties can be calculated efficiently.
In general, Betti curves depend both on the simulation box size and the number of grid points that are used to calculate the density contrast. Thus, we use normalized numbers in units of $[\mathrm{Mpc}^{-3}]$ to make the model predictions. 

\subsection{Persistent homology} \label{persistenthomology}

Persistent homology is an important method in TDA that quantitatively describes features in data across a range of scales. It constructs a concise summary of the shape of a dataset, robust to noise and input perturbations. 
The topology of a field is entirely determined by the maxima, saddles, and minima of the field, with topological changes occurring precisely when the density threshold passes through the value of a singularity according to Morse theory \citep{milnor1963}. 

This process begins with a filtration, a nested sequence of structures built from any well-defined field (e.g., continuous fields or discrete point clouds) by varying the chosen parameter. 
Superlevel and sublevel filtrations are two examples of filtrations based on excursion sets, constructed by traversing the field in opposite directions and selecting the regions that exceed or fall below a certain threshold. In practice, by analyzing the negative of the field, a superlevel filtration becomes equivalent to a sublevel filtration of the inverse field. In $N$ dimensions, Alexander duality provides a further correspondence between the $k$th homology of a manifold and the $(N-k-1)$th cohomology of the complement. This implies that the persistent homology of a superlevel filtration can be obtained by analyzing the sublevel filtration of the inverse field. However, there are subtle differences in practice \citep[see Appendix C of][for an illustrative example in the case of cubical complexes]{Bobrowski2020}.
In this work, we adopt a superlevel filtration, i.e., start at a high density threshold and gradually decrease the density threshold, to keep aligned with previous works \citep[e.g.][]{Wilding_topoweb_2021, jalalikanafi_imprint_2024}. However, since the GUDHI package computes sublevel filtrations, we rely on the correspondence discussed above.

During the filtration process, topological features appear, evolve, and eventually disappear. 
This yields topological quantities that vary as functions of the threshold, $\vartheta = 1+\delta(\boldsymbol{r})$.
To illustrate this process, Fig.~\ref{fig:dens_example} shows the structures identified above the specific thresholds $\vartheta$, alongside their corresponding Betti curves. 
Halos, quantified by $\beta_0$, appear first at the highest density filtration ranges. As the density filtration decreases gradually, halos are connected to filaments and tunnels, reflected in reducing $\beta_0$ and raising $\beta_1$. As the threshold decreases further, filaments are connected to walls enclosing voids, manifested as reducing $\beta_1$ and raising $\beta_2$.

\subsection{Persistence diagrams and Betti curves}

To analyze the multi-scale nature and interactions of topological features in greater detail than Betti curves allow, we represent them using persistence diagrams. These diagrams feature each structure in the ``birth" density and ``death" density coordinates, corresponding to the thresholds at which the feature first appears and at which it subsequently disappears.
The ``persistence" of a feature, thus provides a quantitative measure of its significance across scales. 
All points in a persistence diagram lie above the diagonal line ($\vartheta_{\text{birth}} = \vartheta_{\text{death}}$), since a feature must be born before it can die. In a superlevel filtration, this means the birth density is always higher than the death density.

The persistence diagram preserves the complete birth-death pairing, thereby encoding the entire lifecycle of each topological feature. Meanwhile, Betti curves integrate out the pairing information between birth and death events, retaining only the net balance of features at each density threshold.
This additional information enables persistence diagrams to distinguish between features of different lifetimes that would be indistinguishable in a Betti curve, and to capture the nested hierarchy of features across scales.

\begin{figure*}
    \centering
    \includegraphics[width=0.95\linewidth]{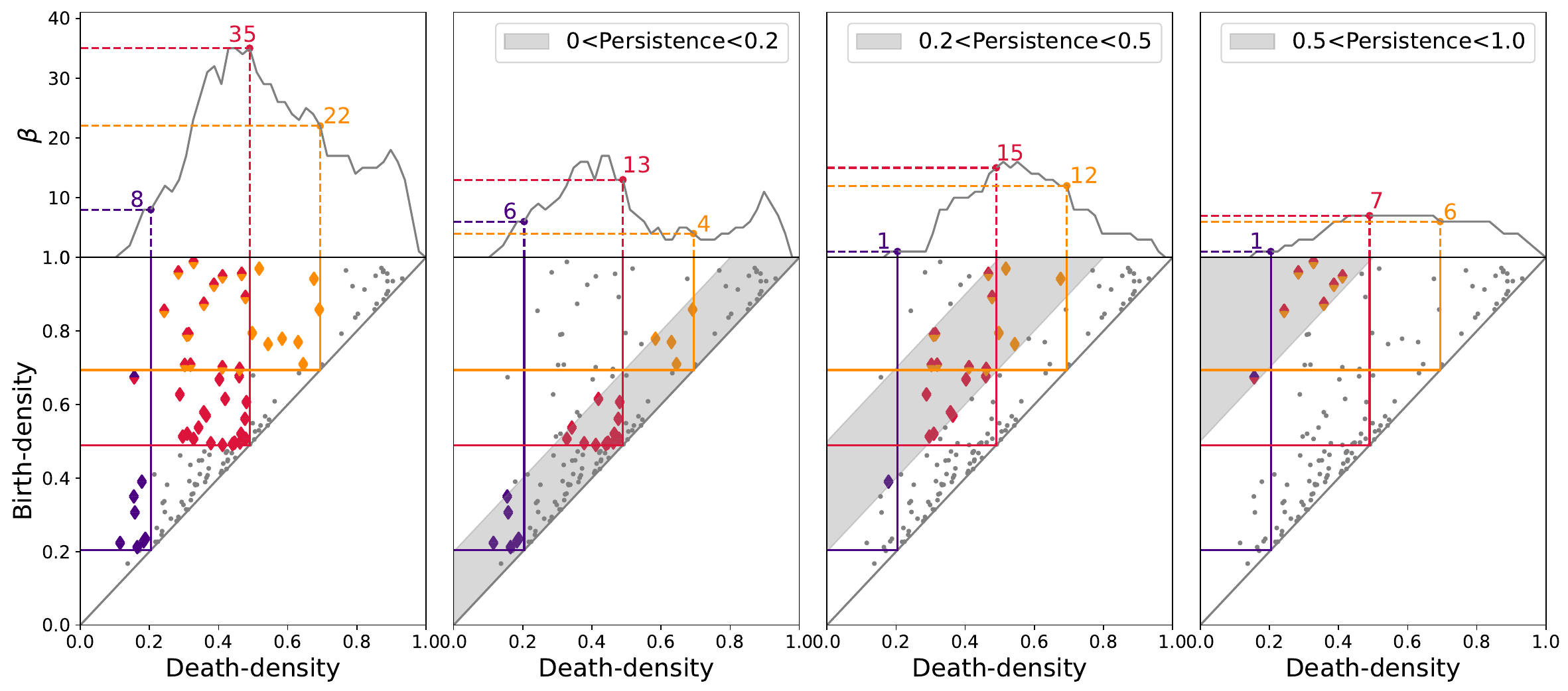}
    \caption{The relationship between Betti curves and persistence diagrams. Upper panel: Betti curves showing the count of topological features as a function of density threshold. The coloured dots on the curves mark the Betti numbers at three specific density thresholds: 0.2 (red), 0.5 (orange), and 0.7 (purple). Bottom panel: The corresponding persistence diagrams, where each point represents a topological feature's birth and death density. The coloured diamonds highlight the specific features that are active (i.e., born before and dying after) at each of the three density thresholds. Features active at multiple thresholds are shown with multiple colours. The three columns on the right demonstrate the effect of applying different persistence cuts (i.e., minimum feature lifetimes) when computing the Betti curves: [0, 0.2], [0.2, 0.5], and [0.5, 1.0], as indicated by the shaded regions in the persistence diagrams.}
    \label{fig:perst}
\end{figure*}

Fig.~\ref{fig:perst} illustrates the relationship between Betti curves (top panel) and persistence diagrams (bottom panel) using simplified mock data. The Betti number $\beta_k(\vartheta)$ for topological features of dimension $k$ is mathematically related to the persistence diagram through the expression,
\begin{equation}
{\beta}_k(\vartheta)=\sum_{i=1}^{n_d} \Theta\left(\vartheta_{i, \text { birth }}^{k}-\vartheta\right) \Theta\left(\vartheta-\vartheta_{i, \text { death }}^{k}\right) ,
\end{equation}
where $\Theta$ denotes the Heaviside step function. This corresponds to counting the number of topological features in the persistence diagram that were born at a density threshold $\vartheta < \vartheta_{i, \text{birth}}$ and die at a threshold $\vartheta > \vartheta_{i, \text {death}}$, that is, points lying to the left of and above the coordinates $(\vartheta_{i, \text {death}}, \vartheta_{i, \text{birth}})$, as shown in Fig.~\ref{fig:perst}. We demonstrate this correspondence explicitly using three representative density thresholds (0.2, 0.5, and 0.7), indicated by red, yellow, and purple highlighting, respectively. Solid lines mark the counting regions, and diamond points represent features counted at each threshold. Points marked with multiple colours correspond to persistent topological features that contribute to the Betti counts across several density thresholds.

Furthermore, the perpendicular distance of a point from the diagonal in a persistence diagram quantifies the persistence (or lifetime) of the corresponding topological feature, defined as,
\begin{equation}
\vartheta_{i, \text { pers }}^{k} \equiv\left|\vartheta_{i, \text { birth }}^{k}-\vartheta_{i, \text { death }}^{k}\right| .
\end{equation}
This measure serves a dual purpose. First, features with short persistence typically originate from noise, offering a natural criterion to distinguish significant structures from random fluctuations. Second, the distribution of feature lifetimes itself encodes valuable physical and cosmological information.
In the three right subpanels of Fig.~\ref{fig:perst}, the grey shaded regions indicate distinct persistence intervals: [0, 0.2], [0.2, 0.5], and [0.5, 1.0]. The Betti curves derived from the lowest persistence range (0–0.2) exhibit high fluctuations, indicative of noise-dominated features. In contrast, Betti curves constructed from features with higher persistence (e.g., 0.2–0.5 or 0.5–1.0) show smoother behaviour and reveal more robust large-scale topological structure. The morphological differences between these filtered Betti curves highlight the multi-scale nature of the cosmic web and demonstrate how persistence-based filtering can isolate physically distinct components of the matter distribution.

\subsection{Persistence strips} 

Since persistence quantifies the lifetime of topological features across filtration scales, it is natural to incorporate it as an additional dimension in the analysis of cosmic web structures. We introduce the concept of persistence strips, defined as follows,
\begin{equation}
    \begin{aligned}
    {\beta}_{k, \pi_p}(\vartheta)=\sum_{i=1}^{n_k} \Theta\left(\vartheta_{i, \text { birth }}^{k}-\vartheta\right) \Theta\left(\vartheta-\vartheta_{i, \text { death }}^{k}\right)
    \\ \Theta\left(\vartheta_{i, \text { pers }}^{k}  - \pi_\text{low,p}\right) \Theta\left(\pi_{\text{high},p} - \vartheta_{i, \text { pers }}^{k}\right) ,
    \end{aligned}
\end{equation}
where $\pi_\text{low,p}$ is the $p^\mathrm{th}$ lower persistence cut and $\pi_\text{high,p}$ the higher cut. These persistence strips can reveal additional information compared to the Betti curves in the following ways. This formulation counts the number of topological features of dimension $k$ that are present at the density threshold $\vartheta$ and whose persistence (lifetime) is in the range $\left[\pi_\text{low,p}, \pi_\text{high,p}\right]$. 

The persistence strips provide a summary that captures both the abundance and robustness of topological features across densities and scales, circumventing the computational cost associated with full two-dimensional image-based approaches.
Furthermore, the persistence (lifespan) of identified features serves as a natural indicator of significance: long-persisting structures generally correspond to physically meaningful signals, while short-lived features can reflect the noise level. This intrinsic property allows for straightforward and physically interpretable filtering based on signal-to-noise ratio, enhancing the robustness of topological summaries in cosmological analysis.

\subsection{Topological structure dependence on neutrino mass}\label{betti_neutrino}
In this subsection, we check the persistent homology of the simulated universe using FLAMINGO simulations. 
We illustrate the basic features and the neutrino imprints on Betti curves and persistence strips for different mass components and different structures. 
We reach two preliminary conclusions. First, both the cold dark matter and all matter fields contain imprints of the neutrinos. Second, among all topological structures, the 2-cycles, which refer to shell-enclosing voids, are most sensitive to neutrino mass.

\begin{figure*}
    \centering
    \includegraphics[width=0.95\linewidth]{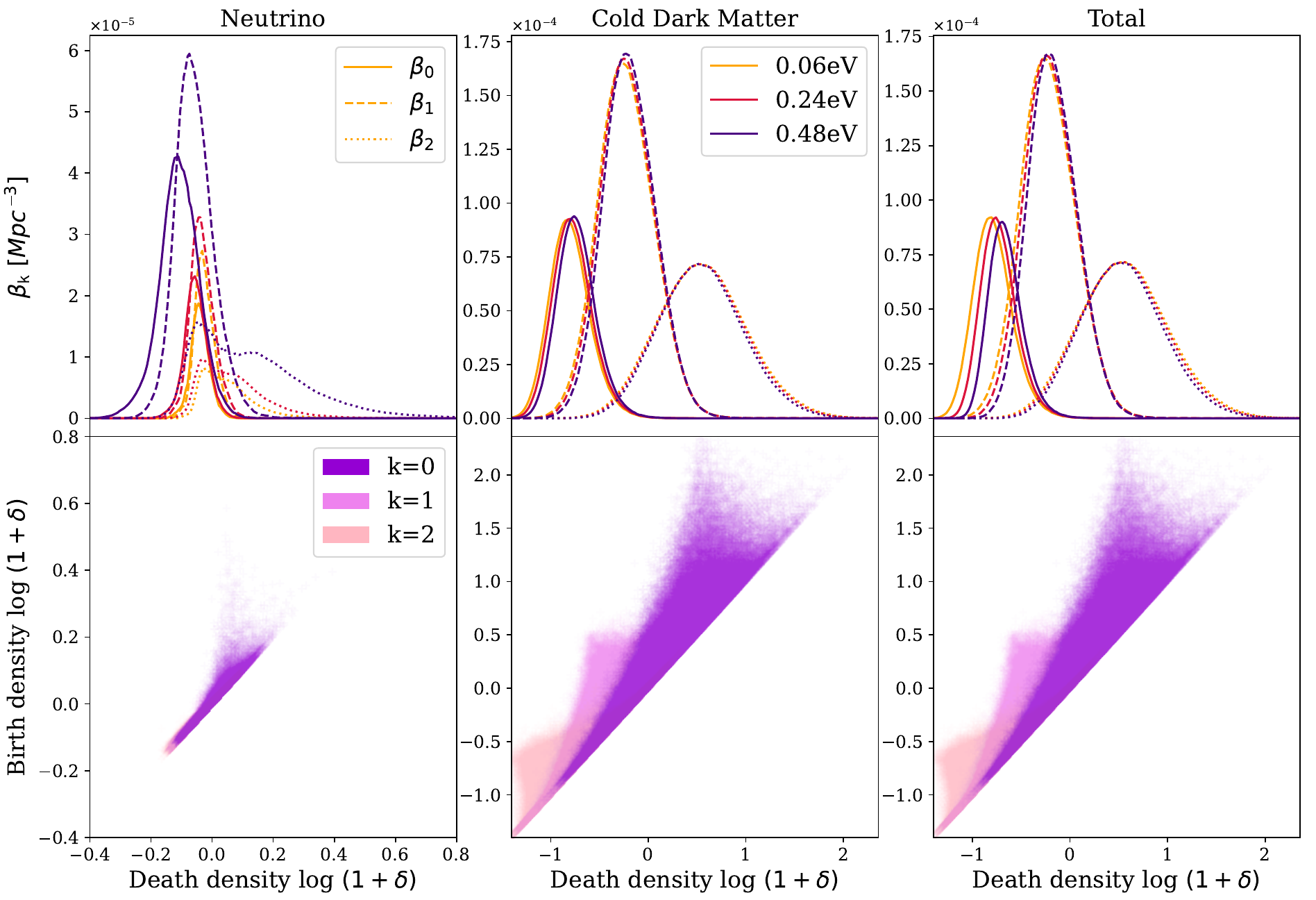}
    \caption{Topological structure of different mass components in the FLAMINGO simulations. Top panel: Betti curves $\beta_k$ for the neutrino, cold dark matter, and total matter density fields. The curves show the count of connected components ($\beta_0$, dotted lines), loops ($\beta_1$, dashed lines), and voids ($\beta_2$, solid lines) as a function of density threshold. Colours indicate the total neutrino mass: 0.06 eV (yellow), 0.24 eV (red), and 0.48 eV (purple). Bottom panel: Persistence diagrams for the 0.06 eV neutrino mass case, corresponding to the density fields shown in the top row. Each point represents a topological feature, with its birth and death coordinates indicating the density thresholds at which it appears and disappears. Structures in dimension $k=0, 1,$ and $2$ are represented in purple, violet, and pink, respectively. This figure illustrates the imprint of neutrino mass on the multiscale topology of cosmic density fields, providing a qualitative basis for comparing the simulations.}
    \label{fig:mul-density}
\end{figure*}

\begin{figure*}
  \centering
  
  \begin{minipage}[b]{0.95\linewidth}
      \centering
      \includegraphics[width=0.95\linewidth]{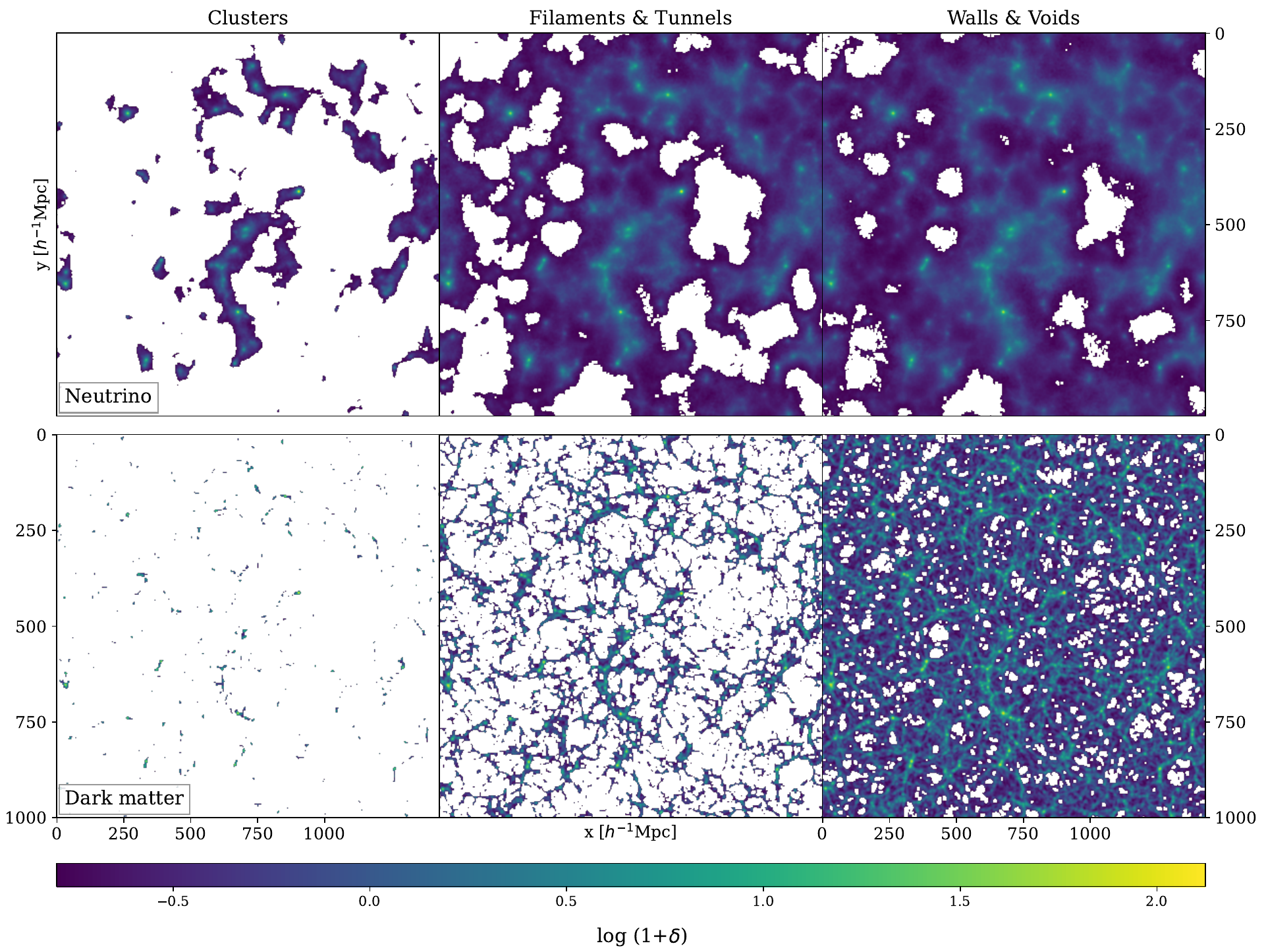}
  \end{minipage}
  \begin{minipage}[b]{0.95\linewidth} 
      \centering
      \includegraphics[width=0.95\linewidth]{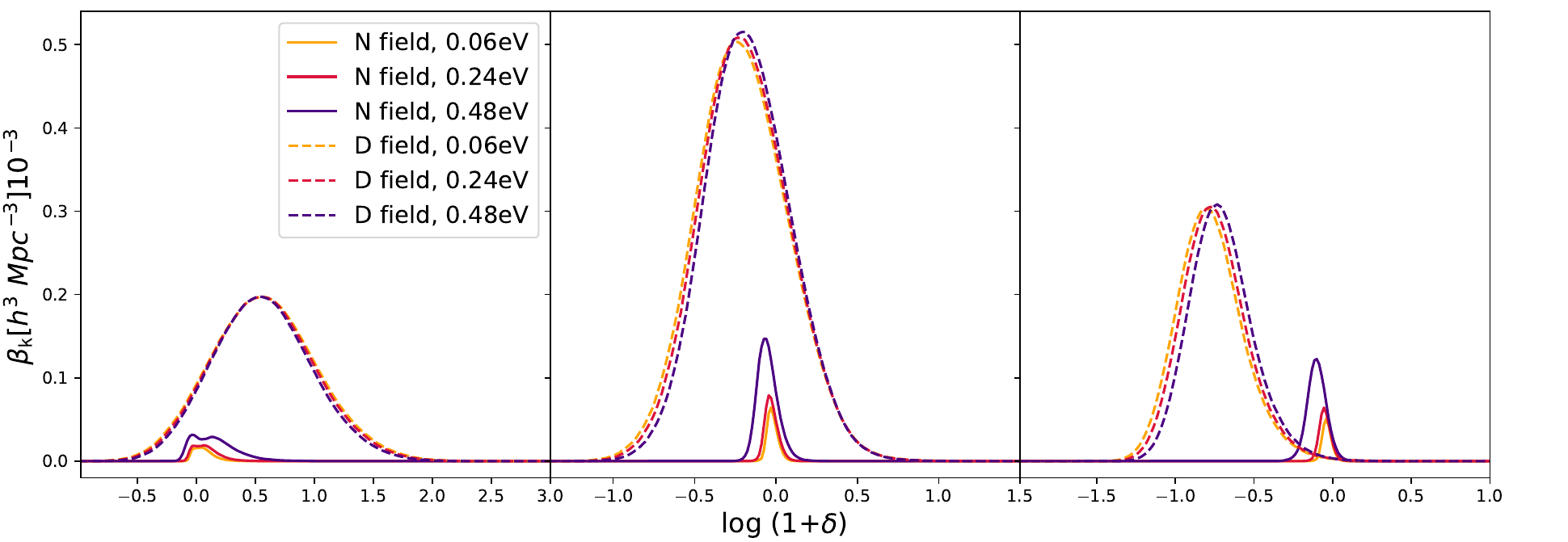}
  \end{minipage}
    \caption{Density superlevel sets and Betti curves of different mass components and structures. Top panel: Superlevel sets of different mass components (neutrinos on the top and dark matter on the bottom) in a 6.8 Mpc slice around a height of 275 Mpc, in the baseline simulation. The structure is depicted as superlevel sets, adopting density thresholds equal to the peak positions of the respective Betti curves to illustrate the disjoint nodes of the cosmic web ($\vartheta =3.53, 1.11$), its filamentary structure ($\vartheta =0.56, 0.93$), and the walls enclosing the cosmic voids ($\vartheta =0.16, 0.91$). Bottom panel: Betti curves as a function of density threshold for connected components ($\beta_0$, left), loops ($\beta_1$, middle), and voids ($\beta_2$, right) of neutrino (solid lines) and cold dark matter (dashed lines).}
    \label{fig:mul-scale}
\end{figure*}

In Fig.~\ref{fig:mul-density}, we show the Betti curves and persistence diagrams built under the superlevel filtration for different matter density fields of FLAMINGO realisations with different neutrino masses, 0.06, 0.24, and 0.48~eV and other cosmological parameters fixed. 
The neutrino field has far fewer structures in total than the cold dark matter field and total matter field. 
Lighter neutrinos further suppress neutrino structures, driving them closer to the mean density. 
In contrast, the topological structures of the cold dark matter field exhibit the opposite response, becoming more pronounced and moving further away from the mean density with lighter neutrino mass. 

Figure~\ref{fig:mul-scale} presents the density superlevel sets and Betti curves of the neutrino field and cold dark matter field. The comparison directly reveals the key feature of the neutrino field relative to cold dark matter: it traces significantly larger scales and thus produces fewer structures overall.
Moreover, voids in the neutrino field occur at markedly different density thresholds than their CDM counterparts. This behaviour directly reflects the formation and evolution of the hierarchical void population: neutrinos give rise to shallow, large-volume voids, in contrast to the intricate, multiscale cold dark matter voids \citep[see e.g., ][]{Sheth2004, Platen2008}.

The total matter field, as the combination of cold dark matter and neutrinos, exhibits Betti curves similar to those of the cold dark matter field. Counterintuitively, however, the differences between neutrino mass scenarios are slightly more pronounced in the total matter Betti curves, particularly in the low-density region where the relative neutrino density is higher and thus their contribution more significant. This arises because the rescaling effect of neutrinos on the density field is substantially stronger than the clustering of neutrinos themselves. This will be further discussed in Section \ref{sec:discuss}. Among all topological structures in the CDM and total matter fields, 2-cycles—which trace the shells enclosing voids—exhibit the greatest sensitivity to neutrino mass. This finding aligns with previous work showing that cosmic voids are more sensitive to neutrino mass than halos, while filaments also serve as a competitive probe.

Unfortunately, we cannot directly detect or even trace the neutrino field in real observation data. Weak lensing observations probe the total matter distribution, including neutrinos, but generally in 2-dimensional projection, while galaxy clustering in redshift space is a biased tracer of the three-dimensional cold matter distribution, which excludes neutrinos to first order. 
Thus, we will focus on the cold dark matter (D) field and total matter (T) field hereafter.

As shown in Fig.~\ref{fig:mul-density}, $\beta_1$ structures (filaments) outnumber both $\beta_0$ (halos) and $\beta_2$ (voids) by roughly a factor of two. This abundance aligns with the function of filaments in connecting halos and surrounding voids.
In our analysis, the total number and peak height of $\beta_2$ features exceed those of $\beta_0$. This contrasts with the findings of \citet{Wilding_topoweb_2021} and \citet{jalalikanafi_imprint_2024}, who report more persistent structures in $k=0$ than in $k=2$. 
We have verified that this is not due to the higher particle resolution, smaller smoothing scale, or differences between the simulations used in our analysis and those used in prior works. 
We obtain similar results when down-sampling the particles to match the resolution of these prior works, when increasing the smoothing scale to 5 Mpc, or when analyzing snapshots from the Quijote simulations instead of FLAMINGO. We speculate that the difference may relate to asymmetry between superlevel and sublevel filtrations (see Section~\ref{persistenthomology}).
Nevertheless, we observe that a large fraction of structures in $k=2$ exhibit low persistence. If we adopt a persistence threshold of 0.2, the number of halos exceeds that of voids.

\begin{figure*}
    \centering
    \includegraphics[width=0.95\linewidth]{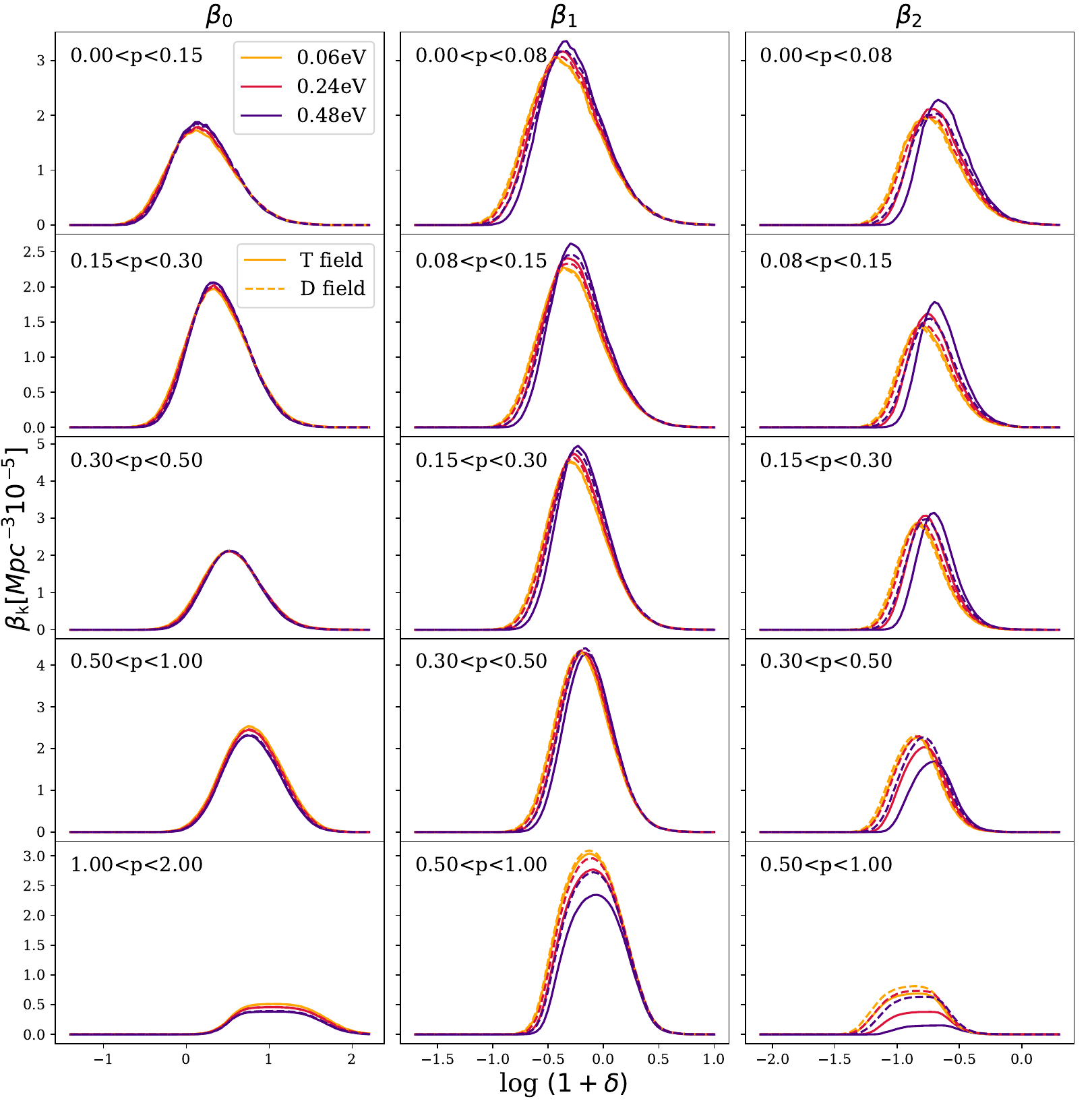}
    \caption{Persistence strips for different mass components in the FLAMINGO simulation, shown for connected components ($k=0$), loops ($k=1$), and voids ($k=2$) from left to right. Each subpanel corresponds to a different persistence bin, as indicated. Solid and dashed lines represent the T field and D field, respectively. Colours indicate different neutrino mass sums: 0.06 eV (red), 0.24 eV (orange), and 0.48 eV (green).}
    \label{fig:sliced}
\end{figure*}

In Fig.~\ref{fig:sliced}, we show the persistence strips of D and T fields across five persistence bins in FLAMINGO realisations with different neutrino masses.
Rather than using identical persistence bins across dimensions, we manually select bin boundaries for each dimension to ensure roughly comparable Betti curve binning. This approach is motivated by Fig.~\ref{fig:mul-density}, which shows that topological features of different dimensions exhibit distinct persistence ranges, with $k=2$ structures displaying the largest spread.
Both the D and T fields exhibit a clear capacity to distinguish neutrino masses with filamentary and void structures (topological dimensions $k=1,2$) across all persistence thresholds $\pi_p$, with the T field generally providing higher discriminatory power.
As persistence increases, the peak positions of $\beta_{0, \pi_p}$ and $\beta_{1, \pi_p}$ shift slightly toward higher densities, whereas $\beta_{2, \pi_p}$ peaks at lower densities. This indicates that longer-lived halos and filaments are associated with higher-density regions, while long-lived voids reside in lower-density environments.

Neutrino mass significantly influences the amplitude and shape of the topological distributions. 
Higher neutrino mass pushes the peak closer to the mean density across all three topological dimensions, as is the case for Betti curves.
In low-persistence bins, higher neutrino mass leads to an increased number of structures. 
However, this trend reverses at high persistence: the amplitude decreases with increasing neutrino mass, suggesting that massive neutrinos suppress the formation of long-lived structures. 
A transition region between these regimes shows minimal sensitivity to neutrino mass. Notably, the relative difference in amplitude between neutrino masses is largest in the highest persistence bin.

In low-persistence bins, an increase in neutrino mass produces a tighter distribution of topological features. This `tightening' pattern signifies that massive neutrinos compress the evolutionary timeline of transient structures: they cause structures to emerge at lower density thresholds and dissolve at higher densities during the filtration process. This behaviour implies a sharply peaked dissolution rate, which rises rapidly to a maximum before declining steeply. In contrast, high-persistence structures exhibit a flatter distribution across densities, reflecting a more uniform dissolution rate and highlighting the distinct evolutionary behaviour of structures with varying persistence.

\section{Emulation and cosmological parameter extraction}\label{sec:emu}

To evaluate the sensitivity of Betti curves to cosmological parameters and assess their capacity to break degeneracies among them, we construct an emulator based on Gaussian process  regression across a 10-dimensional cosmological parameter space. This emulator enables efficient interpolation of Betti curves as functions of cosmological parameters and density thresholds, supporting subsequent cosmological inference through Bayesian posterior sampling and evidence estimation.

\subsection{Emulator construction and training}

Gaussian process (GP hereafter) provide a non-parametric, probabilistic framework for approximating smooth functions from finite training data. A GP defines a distribution of functions, characterized by a mean function $\mu(\theta)$ and a covariance kernel $k(\theta, \theta')$, which models the similarity between the input points $\theta$ and $\theta'$. Given a training set of cosmological simulations with parameter sets $\left\{ {\theta} \right\}_j$, where $j$ refers to the training samples index, the GP predicts the function value of persistence strips $ {\beta_{d,\pi_p}(\vartheta; \left\{ {\theta} \right\}_j)}$ at a new point $\left\{ {\theta} \right\}_*$ as a Gaussian distribution, 
\begin{equation}
{\beta_{d,\pi_p}(\vartheta; \left\{ {\theta} \right\}_*)} \sim 
\mathcal{G} \mathcal{P}
\left(\mu\left({\theta}_*\right), k\left({\theta}, {\theta}_*  \right)+\sigma_n^2  I\right) ,
\end{equation}
where $\sigma_n^2$ accounts for observational and modelling noise. For a more detailed discussion of GP and their features, see \cite{GPforML_2006}. 

We model each persistence strip across death density thresholds $\vartheta$ using an independent GP. To incorporate the density dimension, we treat $\vartheta$ as an additional input parameter to the 10 cosmological parameters, allowing the emulator to smoothly interpolate topological measurements across both cosmological and density spaces. The full emulator comprises 15 independent 11-dimensional GPs (one for each of the three topology dimensions and five persistence cuts), each yielding predictions with quantified uncertainties.

The choice of representation for the target observable significantly influences emulation accuracy. For instance, the CosmicEmu \citep{Lawrence_CosmicEmu_2010} represents the matter power spectrum as $k^{3/2}P(k)$, and the EuclidEmulator \citep{EUCLIDEMULATOR_2019} uses the nonlinear boost $B(k) = P(k,z)/P_{\text{linear}}(k,z)$ to improve the emulator performance. In this work, we introduce a cumulative representation of the persistence strips to enhance smoothness (especially for the lower density slope) and improve emulation efficacy. The cumulative statistic is defined as:
\begin{equation}
\operatorname{CUM_{k,\pi_p}}(x)=\log _{10}\left(\sum_{n=x}^{\infty} \beta_{k,\pi_p}( \vartheta_n)+C\right) ,
\end{equation}
where $C = 500$ is a constant added to avoid logarithmic numerical instabilities. This transformation significantly improves the smoothness and differentiability of the target function, leading to more robust GP interpolation. 
\begin{figure}
    \centering
    \includegraphics[width=1\linewidth]{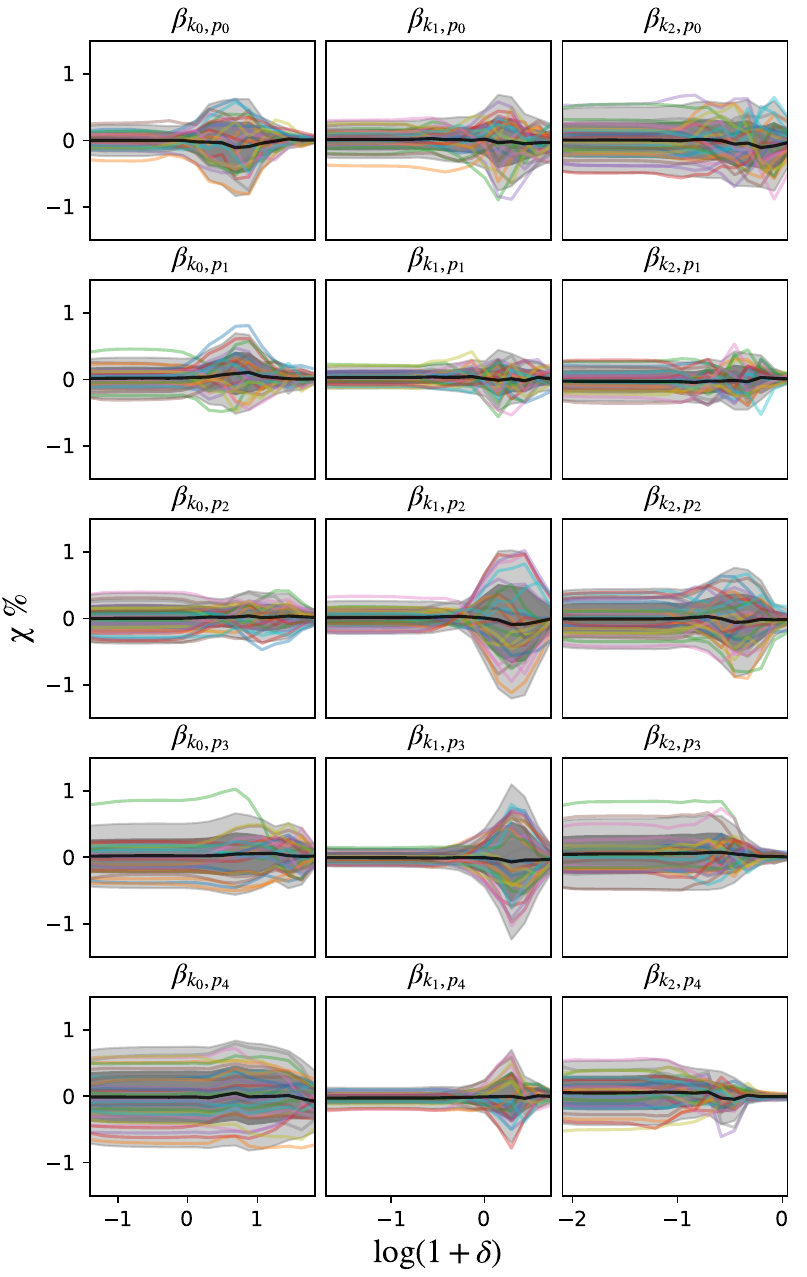}
    \caption{The emulator accuracy of selected testing samples. Different panels show results for different dimensions and different persistence ranges as indicated on top of each panel. Each coloured line represents one testing sample, and the black solid line shows the average error between the prediction and the truth. The dark and light grey shaded regions show the $68\%$ and $95\%$ confidence intervals, respectively. }
    \label{fig:cumsliced}
\end{figure}

For the Gaussian kernel choice, we adopt a composite kernel $k\left(x, x^{\prime}\right)=k_{\text {constant }}+k_{\text {Matern32 }}\left(x, x^{\prime}\right)$ by combining a constant kernel and the $\rm Mat\acute{e}rn32$ kernel,
\begin{equation}
k_{\rm Mat\acute{e}rn32}\left(r^2\right)=\sigma_f^2 \left(1+\sqrt{3 r^2}\right) \exp \left(-\sqrt{3 r^2}\right) 
\end{equation}
where $r = (x-x') / l$, $l$ is the characteristic length scale governing the smoothness of the function, $\sigma_f^2$ represents the signal variance, and $r$ denotes the scaled Euclidean distance between input points.
The constant kernel captures global, non-zero mean trends in the function, while the $\rm Mat\acute{e}rn32$ kernel models smooth, but not infinitely differentiable, local variations. 
The hyperparameters of the kernel, including $\sigma_f$ and $l$, are optimized by maximizing the log marginal likelihood,
\begin{equation}
\ln p(\mathbf{y} \mid X)=-\frac{1}{2} \mathbf{y}^{\top} K^{-1} \mathbf{y}-\frac{1}{2} \ln |K|-\frac{n}{2} \ln 2 \pi
\end{equation}
where $K=k\left(x, x^{\prime}\right)+\sigma_n^2 I$ is the covariance matrix of the observed data, $K(X, X)$ is the kernel matrix evaluated at the training points, $\sigma_n^2$ is the noise variance, and $\mathbf{y}$ is the vector of observed function values.

In this work, we employ the python code george3 developed by \citet{hodlr} \footnote{http://george.readthedocs.io/en/latest/} to develop our emulator. The emulator is constructed based on the 50 simulations as outlined in Section~\ref{sec:data_emu}, where 40 simulations are used as the primary training sample, with the remaining 20\% of simulations held out as the testing set to validate performance. To ensure robustness against specific random subdivisions, this train-test partitioning process was repeated multiple times with different random seeds. 
The cosmological parameter space explored in this work is defined by the 10-dimensional Latin hypercube design detailed in the top panel of Table~\ref{tab:para-Latin}. In addition to this baseline parameterization, we systematically tested alternative parameterizations by substituting $\ln A_s$ with $\sigma_8$.
A comparative analysis of emulator performance under the different parameterization is provided in Appendix \ref{app:emulator}. We find $\sigma_8$ yields much higher emulation accuracy to $\ln A_s$, indicating its strong correspondence with topological features of the cosmic web.

\subsection{Emulator accuracy}
Performance was assessed using an iterative holdout validation: in each iteration, a random 20\% of the simulations were withheld as a test set, and the emulator was retrained on the remaining 80\%. This process was repeated over 10 iterations to ensure statistical robustness.
The accuracy of the emulator was quantified using the relative residual of the cumulative statistic:
\begin{equation}
\chi_{\mathrm{CUM_{k,\pi_p}}{\left( x \right) }} = 
\frac{\mathrm{CUM_{k,\pi_p}}\left( x \right) - \langle \mathrm{CUM_{k,\pi_p}}\left(x \right) \rangle_{j}}{\langle \mathrm{CUM_{k,\pi_p}}\left( x \right) \rangle_{j} }
\end{equation}
where $j$ represent the repeating iterations and $x$ denotes density threshold used for cumulating. Our analysis focuses on the accuracy of the cumulative persistence strips, as this is the format directly utilized in our cosmological parameter inference. We have examined the residuals of the non-cumulative (original) format, which, as expected, exhibit numerical instabilities, particularly at the edges of the curves.

Fig.~\ref{fig:cumsliced} displays the corresponding residuals, with individual realisations shown as coloured lines. Dark and light shaded regions indicate the 68\% and 95\% confidence intervals, respectively, across all test samples. The emulator achieves high accuracy, with residuals $\leq 1\%$ generally. For the structures in the highest persistence bin or with $k=2$, the error is relatively higher, reflecting the greater dynamic range and structural complexity in those regimes. These results demonstrate the emulator’s reliability across a wide range of topological scales and cosmological models.

\subsection{Bayesian posterior and evidence estimation}

For cosmological parameter inference, we employ the Nautilus sampler \citep{nautilus}\footnote{https://nautilus-sampler.readthedocs.io/en/stable/}, a Bayesian nested sampling algorithm that utilizes neural network-based importance sampling to significantly reduce the number of required likelihood evaluations. This approach efficiently estimates both posterior distributions and Bayesian evidence while incorporating all likelihood evaluations during the sampling process. We adopt flat priors on all cosmological parameters, with ranges identical to those used in the Latin hypercube design of our training simulations. For this analysis, we find that 6,000 live points provide sufficient sampling accuracy and stability.

The covariance matrix used in the likelihood evaluation combines contributions from both cosmic variance and emulator prediction errors. Cosmic variance is estimated via bootstrap resampling of the FLAMINGO $2.8,\mathrm{Gpc}$ simulation box. The simulation, configured with the same resolution and smoothing scales as the emulator, is divided into 512 sub-boxes of $350,\mathrm{Mpc}^3$ each, from which the covariance is robustly estimated.
The emulator uncertainty is incorporated as a diagonal matrix, where each diagonal element corresponds to the Gaussian process prediction error in the respective density bin.

To incorporate limited cosmological information from the CMB, we add the CMB prior term to the total likelihood function following \citet{DESI-DR2cosmology2025}, 
\begin{equation}
\chi_{\rm CMB}^2=\left(\begin{array}{c}
\theta_{*, 100}-\theta_{*, 100}^{\text {true }} \\
\omega_{\mathrm{cdm}}-\omega_{\mathrm{cdm}}^{\text {true }}\\
\omega_{\mathrm{b}}-\omega_{\mathrm{b}}^{\text {true }} 
\end{array}\right)^{\mathrm{T}} C^{-1}\left(\begin{array}{c}
\theta_{*, 100}-\theta_{*, 100}^{\text {true }} \\
\omega_{\mathrm{cdm}}-\omega_{\mathrm{cdm}}^{\text {true }}\\
\omega_{\mathrm{b}}-\omega_{\mathrm{b}}^{\text {true }} 
\end{array}\right) ,
\end{equation}
where $\omega_{\mathrm{b}} \equiv \Omega_{\mathrm{b}} h^2$, $\omega_{\mathrm{cdm}} \equiv \Omega_{\mathrm{cdm}}  h^2$, 
and $\theta_{*, 100}$ is 100 times the angular acoustic scale of the CMB power spectrum, defined as $\theta_*=r_* / D_M\left(z_*\right)$, where $r_*$ is the comoving sound horizon at recombination and $D_M\left(z_*\right)$ is the comoving angular diameter distance to the last scattering surface at redshift $z_*$ \citep{Lemos_CMB_2023}. The true value of these parameters is calculated from the cosmology of the simulation to be estimated. The covariance matrix $C$ of these three parameters is derived conservatively from \emph{Planck} and ACT data, following \citet{Elbers25_DESI}.
This term ensures consistency between our inferred parameters and early-universe constraints, thereby enhancing the robustness of the cosmological inference.

\section{Results}\label{sec:perfm}

In this section, we present an analysis of the cosmological dependence captured by our emulator and discuss the resulting cosmological constraints derived from these topological summaries. 
We examine how this dependence varies across different cosmic environments and   investigate the cosmological inference for key parameters and from different probes and different fields.

\begin{figure*}
    \centering
    \includegraphics[width=0.95\linewidth]{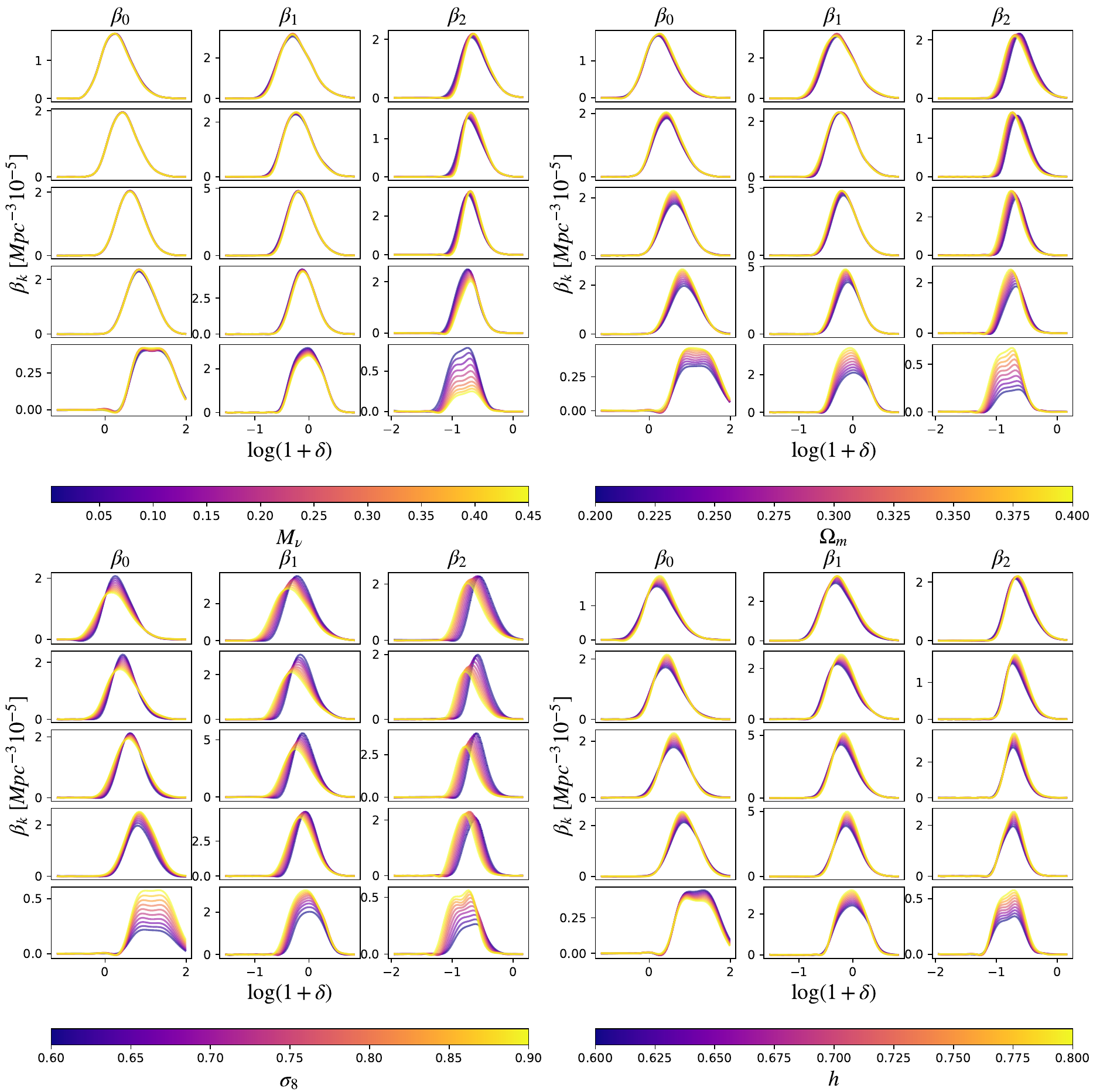}
    \caption{Dependence of persistence strips on key cosmological parameters, including the neutrino mass $\SigmaMnu$ (upper left), the matter density $\Omega_\mathrm{m}$ (upper right), the amplitude of matter fluctuations $\sigma_8$ (bottom left), and the Hubble constant $h$ (bottom right). All other parameters are fixed at the centre of the hypercube. In each panel, the five rows correspond to the five persistence ranges as in Fig.~\ref{fig:sliced}, illustrating how the sensitivity to each parameter varies with topological persistence. Colours in each panel represent the varying value of the parameter of interest, as indicated by the colour bar.}
    \label{fig:dependence}
\end{figure*}

\subsection{Cosmological parameter dependence}\label{cosmodepend}
Among the ten cosmological parameters considered, we find that the neutrino mass sum $\SigmaMnu$, the matter density $\Omegam$, the amplitude of matter fluctuations $\sigma_8$, and the Hubble constant $h$ exert the most pronounced influence on the topology of the cosmic web, as quantified by the Betti curves.
As illustrated in Fig.~\ref{fig:dependence}, each parameter affects the Betti curves in a distinct way, with other parameters fixed at the centre of the hypercube. Persistence strips shown in each row correspond to different persistence ranges, as defined in Fig.~\ref{fig:sliced}.

The topological statistics of clusters ($k=0$), displayed in the first column of each panel in Fig.~\ref{fig:dependence}, exhibit a comparatively modest response to variations in cosmological parameters. The most substantial influences arise from $\Omegam$ and $\sigma_8$, followed by a weaker dependence on $h$; the response to $\SigmaMnu$ is nearly negligible (at fixed $\Omega_\mathrm{m}$ and $\sigma_8$).
The matter density $\Omega_{\mathrm{m}}$ systematically enhances the abundance of cluster structures across all persistence levels, with a magnitude that scales with the persistence value. In contrast, the amplitude of matter fluctuations $\sigma_8$ exerts a more complex influence: it suppresses the number of low-persistence clusters while enhancing the population of high-persistence ones. This redistribution consequently shifts the peak of the low-persistence distribution to lower density thresholds. The Hubble parameter $h$ exhibits the converse trend to $\sigma_8$; an increase in $h$ shifts structures from high to low persistence bins and moves the peak of the low-persistence distribution toward higher densities.
In the case of neutrino mass, the $k=0$ response is too small to be statistically significant.

Filamentary structures ($k=1$), as illustrated in the second column of each panel in Fig.~\ref{fig:dependence}, generally encode richer cosmological information than $k=0$. Among the four cosmological parameters studied, $\sigma_8$ exhibits the most pronounced influence, followed by $\Omegam$, with $h$ and $\SigmaMnu$ showing comparatively weaker but still discernible effects.
An increase in neutrino mass $\SigmaMnu$ suppresses the abundance of high-persistence filaments while enhancing low-persistence ones, accompanied by a systematic peak shift toward higher density thresholds. In contrast, a higher matter density $\Omegam$ primarily boosts the population of high-persistence filaments and shifts the distribution peak toward lower densities. The amplitude of matter fluctuations $\sigma_8$ displays a trend consistent with that observed in the cluster environment, promoting the transition of structures from low to high persistence bins. Similarly, a larger Hubble parameter $h$ results in a sharper filament distribution across all persistence levels and an increase in the number of high-persistence filaments.

The topological statistics of cosmic voids ($k=2$), presented in the right-most column of each panel in Fig.~\ref{fig:dependence}, exhibit the most complex and informative behaviour among all topological components, showing strong dependence on all four cosmological parameters.
The influence of neutrino mass is particularly striking, characterized by a combined ``shifting" and ``narrowing" pattern across persistence bins. A higher $\SigmaMnu$ increases the abundance of low-persistence voids and shifts their distribution toward higher density thresholds, while simultaneously suppressing the number of high-persistence voids. This can be understood as a topological manifestation of neutrino free-streaming: the diffusive effect of heavier neutrinos elevates the density floor in deep potential wells, effectively fragmenting large, persistent voids into smaller, shallower ones.
Similarly, the matter density $\Omegam$ exerts a strong influence, significantly enhancing the number of high-persistence voids while leaving low-persistence counts largely unaffected. This is accompanied by a systematic shift of the entire distribution toward lower density thresholds. The impact of $\sigma_8$ on void statistics follows a similar pattern with that observed in filaments but with slightly larger variance. The Hubble parameter $h$ also shows a trend analogous to its effect in the filamentary environment.

Several cosmological parameters, including the baryon fraction $\fb$, the scalar spectral index $\ns$, the running of the spectral index $\rm \alpha_s$, exhibit only a minor influence on the Betti curves. Similarly, parameters associated with decaying dark matter ($\Gammadcdm$) and standard dark energy evolution ($w_0$, $w_a$) show negligible impact across all topological features. A comprehensive presentation of the effects of these parameters is provided in Appendix~\ref{app:cosmo}. Due to the insensitivity of most of the parameters, we will focus on the four main parameters in the following analysis.

It is essential to leverage the distinct responses of topological components to break degeneracies among cosmological parameters. For instance, filamentary structures ($\beta_1$) respond in opposite directions to changes in $\SigmaMnu$ and $\sigma_8$, suggesting a potential degeneracy between neutrino mass and matter clustering amplitude. The degeneracy can be effectively broken by leveraging the complementary parameter sensitivities of different topological environments, namely, by jointly exploiting the clustered (cluster, $k=0$) and underdense (void, $k=2$) structures. This multi-scale approach enhances the overall constraining power and reduces parameter uncertainties.

\subsection{Cosmological parameters inference}

In this subsection, we first analyze the four parameters of primary interest ($\Omega_{\mathrm{m}}$, $\h$, $\rm \sigma_8$, $\SigmaMnu$), then compare the constraints on all parameters obtained with and without CMB priors. 
It should be noted that no observational noise has been added in this section. Consequently, the accuracy of the recovered cosmological parameters is limited by both the emulator and the construction of the covariance matrix, which is based on a volume of $(350\,\mathrm{Mpc})^3$.
We randomly select one of the testing samples not used in the emulator training as the example. The inferred inference of the four primary cosmological parameter are listed in Table~\ref{tab:bb_compare}. 
\begin{figure}
    \centering
    \includegraphics[width=1\linewidth]{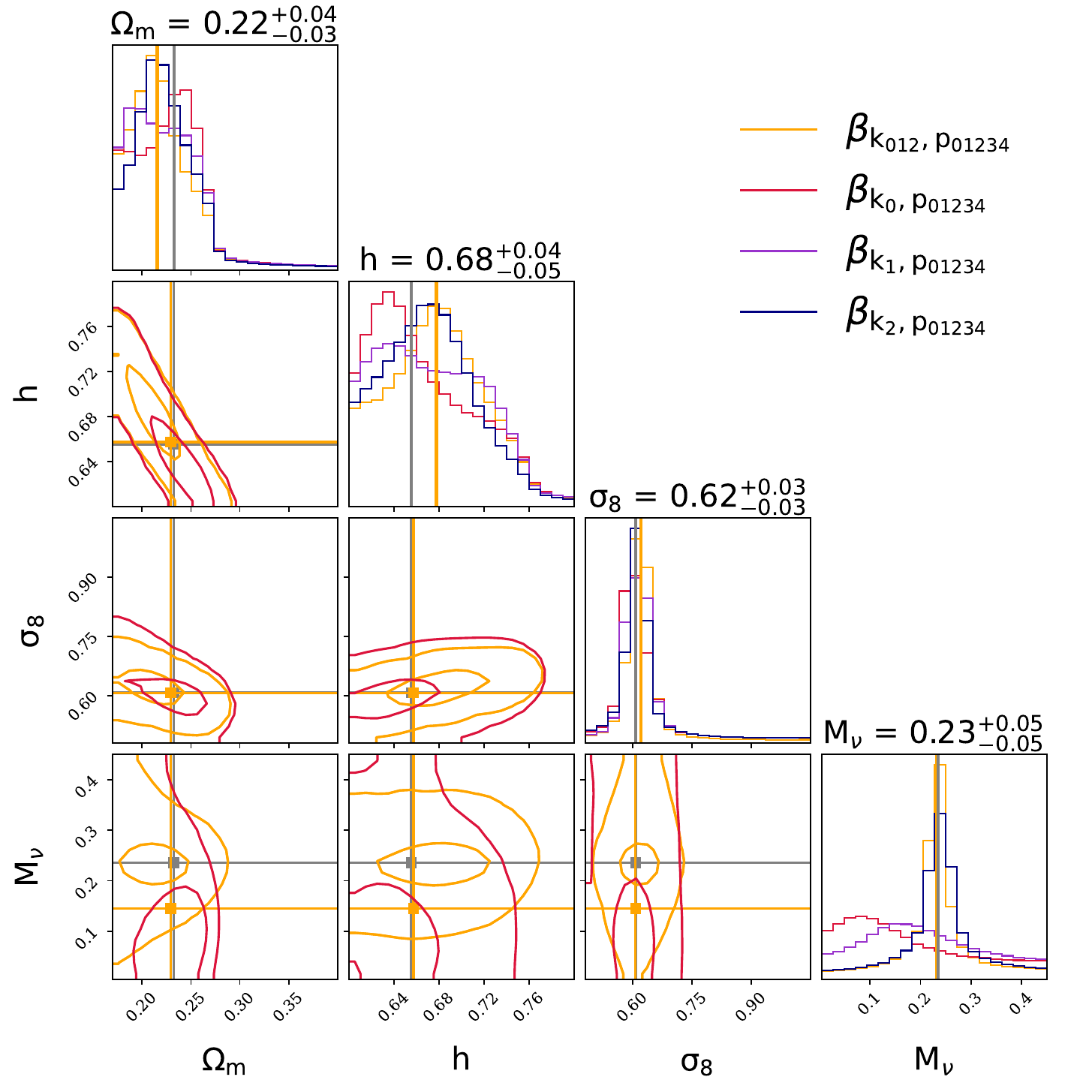}
    \caption{Cosmological constraints from persistence analysis. Confidence contours for $\Omegam$, $\h$, $\rm \sigma_8$, and $\SigmaMnu$ using intermediate persistence strips across dimensions $k=0,1,2$ individually and jointly with CMB priors. For clarity, we only show the $k=0$ (red contour) and joint analysis (orange contour) cases in 2-D subpanels, with true values in grey and the median value of the joint analysis in orange. We also include the results from $k=1$ (purple) and $k=2$ (navy) in the 1-D subpanels. Labels above each column indicate the median and 68\% credible intervals for $\beta_{k_{012},p_{01234}}$.}
    \label{fig:corner_bb_compare}
\end{figure}

\begin{table*}
    \centering
\begingroup
\renewcommand{\arraystretch}{1.3} 

    \begin{tabular}{c |c | c c c c}
    \hline\hline
    Probe &  CMB &   $\Omegam$ & $\h$ & $\rm \sigma_8$ & $\SigmaMnu$/eV \\
& priors &  0.233 & 0.655 & 0.607 & 0.236 \\
    \hline \hline
    $\beta_{k_{0},p_{01234}}$ &
    \checkmark & 
    $0.229_{-0.039}^{+0.029} $ & 
    $0.657_{-0.036}^{+0.063} $ & 
    $0.607_{-0.032}^{+0.069} $ & 
    $0.145_{-0.092}^{+0.174} $ 
    \\
    $\beta_{k_{1},p_{01234}}$ &
    \checkmark & 
    $0.220_{-0.033}^{+0.037} $ & 
    $0.671_{-0.046}^{+0.054} $ & 
    $0.618_{-0.036}^{+0.058} $ & 
    $0.195_{-0.105}^{+0.141} $ 
    \\
    $\beta_{k_{2},p_{01234}}$ &
    \checkmark & 
    $0.221_{-0.026}^{+0.033} $ & 
    $0.670_{-0.041}^{+0.042} $ & 
    $0.615_{-0.035}^{+0.058} $ & 
    $0.238_{-0.061}^{+0.057} $ 
    \\
    \hline
    $\beta_{k_{0},p_{123}}$ &
    \checkmark & 
    $0.229_{-0.041}^{+0.031} $ & 
    $0.658_{-0.038}^{+0.065} $ & 
    $0.607_{-0.034}^{+0.095} $ & 
    $0.146_{-0.100}^{+0.177} $  
    \\
    $\beta_{k_{1},p_{123}}$ &
    \checkmark & 
    $0.212_{-0.025}^{+0.038} $ & 
    $0.683_{-0.050}^{+0.044} $ & 
    $0.625_{-0.039}^{+0.070} $ & 
    $0.183_{-0.093}^{+0.131} $  
    \\
    $\beta_{k_{2},p_{123}}$ &
    \checkmark & 
    $0.218_{-0.031}^{+0.039} $ & 
    $0.675_{-0.048}^{+0.052} $ & 
    $0.613_{-0.041}^{+0.106} $ & 
    $0.228_{-0.097}^{+0.100} $ 
    \\
    \hline
    $\beta_{k_{{012}},p_{123}}$ &
    \checkmark & 
    $0.226_{-0.038}^{+0.033} $ & 
    $0.663_{-0.041}^{+0.060} $ & 
    $0.612_{-0.029}^{+0.053} $ & 
    $0.234_{-0.072}^{+0.059} $  
    \\
    $\beta_{k_{{012}},p_{01234}}$ &
    \checkmark & 
    $0.216_{-0.027}^{+0.036} $ & 
    $0.678_{-0.045}^{+0.044} $ & 
    $0.620_{-0.032}^{+0.035} $ & 
    $0.233_{-0.050}^{+0.049} $  
    \\
    $\beta_{k_{{012}},p_{123}}$ &
       & 
    $0.227_{-0.037}^{+0.046} $ & 
    $0.682_{-0.068}^{+0.080} $ & 
    $0.616_{-0.037}^{+0.042} $ & 
    $0.252_{-0.063}^{+0.072} $ 
    \\
    $\beta_{k_{{012}},p_{01234}}$ &
       & 
    $0.230_{-0.033}^{+0.045} $ & 
    $0.684_{-0.073}^{+0.072} $ & 
    $0.617_{-0.039}^{+0.035} $ & 
    $0.254_{-0.046}^{+0.057} $
    \\
    \hline\hline
    \end{tabular}
    \endgroup
    \caption{Constraints on key cosmological parameters from topological persistence analysis. Median values with 68\% credible intervals for $\Omegam$, $\h$, $\sigma_8$, and $\SigmaMnu$ are presented for a range of analyses, including different topological dimensions ($k$), persistence bins ($\pi_p$), density fields, and data combinations (with/without CMB priors), all with 2 Mpc smoothing. The true parameter values are shown at the top of the table.}
    \label{tab:bb_compare}
\end{table*}

\subsubsection{Four most influential parameters}

We first focus on the posteriors of $\Omegam, \h,\, \sigma_8$, and $\SigmaMnu$ based on the topological measurement from different dimensional probes. 
The constraints separately from each dimension $k$ and jointly are presented in Fig.~\ref{fig:corner_bb_compare}, respectively. For clarity, we only show the $k=0$ and joint analysis cases in 2-D subpanels, with true values in grey and the median value of the joint analysis in orange. We include the results from $k=1$ and $k=2$ for comparison in the 1-D marginalized distributions. 

Consistent with the dependence analysis in Section~\ref{cosmodepend}, all persistence strips individually provide different levels of constraining power on these four parameters. 
In general, $k=2$ provides the tightest constraints within all dimensions, and combining persistence strips in all three dimensions yields tighter constraints.

For $\Omegam$ and $\sigma_8$, the constraining power is comparably good across all three-dimensional topological components, showing no significant differences beyond the $1\sigma$ level.
Combining three-dimensional topological measurements tightens the constraints, indicating that the information on the matter density is encoded across multiple topological dimensions and is complementary to each other. 
For $\h$, the constraints are slightly looser, as predicted in Sec.~\ref{cosmodepend}.
The degeneracy between $\Omega_{\mathrm{m}}$ and $h$ persists across all probes and proves difficult to break using topological information alone. This can be understood from Fig.~\ref{fig:dependence}, which shows that their dependence patterns are nearly identical.
These results partially contrast with previous studies: \citet{Ouellette_betti_kNN_2025} reported that $\Omegam$ and $\rm \sigma_8$ are constrained primarily by $\beta_0$ and $\beta_1$, with little contribution from $\beta_2$, while \citet{calles_cosmology_2024} found clusters and voids to be most informative for $\Omegam$. Our analysis indicates that all three environments contribute significantly to constraining both $\Omegam$ and $\rm \sigma_8$.

\begin{figure}
    \centering
    \includegraphics[width=1\linewidth]{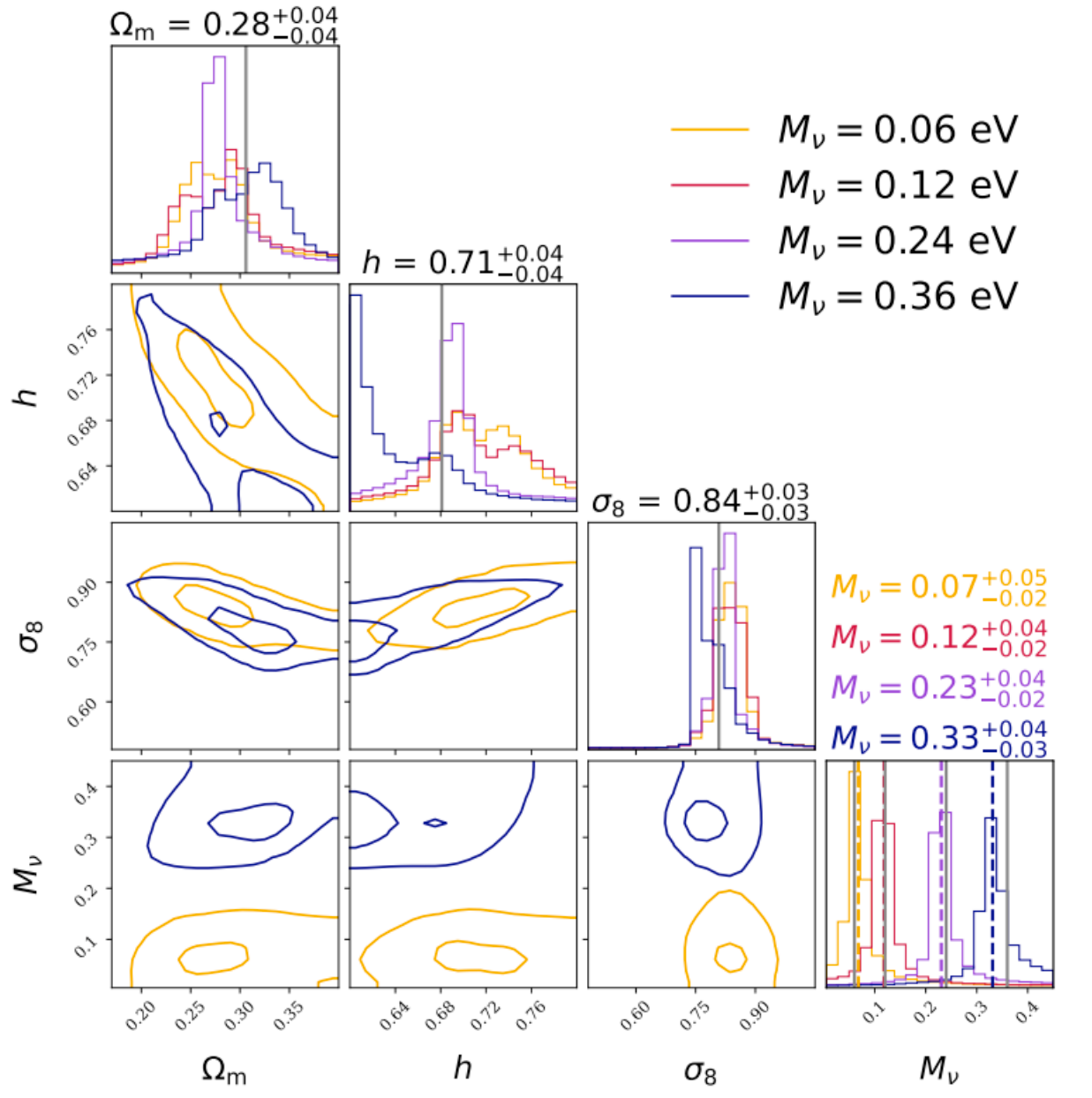}
    \caption{Similar to Fig.~\ref{fig:corner_bb_compare}, but showing the cosmological inference dependence on the true neutrino mass. For clarity, we only show the smallest and largest neutrino mass cases in the 2-D subpanels (in orange and blue, respectively). We include the results from all four cases in the 1-D subpanels, with true values in grey and the median value of each analysis in the corresponding colour. Labels above each column show the median and 68\% credible interval: for the left three columns, this is the 0.06~eV case; for the rightmost column, it includes all four cases, with the colour the same as the corresponding lines.}
    \label{fig:corner_Mnu_dependence}
\end{figure}

The constraints on the neutrino mass $\SigmaMnu$ are dominated by voids ($k=2$). Halos ($k=0$) contribute negligible information, while filaments ($k=1$) offer only modest improvements. 
Voids give 5 times tighter constraints on $\SigmaMnu$ than halos (as shown in Table~\ref{tab:bb_compare}). Combining the three-dimensional probes will tighten the constraints further to an uncertainty of $0.05$~eV.
A notable degeneracy between $\sigma_8$ and $\SigmaMnu$ is evident when using halos only ($k=0$), while the degeneracy is effectively broken upon the inclusion of the void-based measurement, underscoring the value of multi-dimensional topological probes in breaking the parameter degeneracies and enhancing the robustness of cosmological inference.

In Fig.~\ref{fig:corner_Mnu_dependence}, we assess the dependence of constraining power on the true neutrino mass using four additional simulations in which only the neutrino mass is varied. The constraining power remains generally stable, with slightly looser constraints near the edges of the hypercube, which is expected. Other parameters exhibit consistent performance, confirming the robustness of our results. Given this validation and to maintain brevity, we omit results from other test samples or random seeds used in emulator training.
We note that here the constraint on neutrino mass is slightly tighter than in the case shown above. This may be because the other parameters lie closer to the centre of the hypercube, where emulator performance is typically better.

We identify three primary factors that likely contribute to the discrepancies between our constraints and those of \citet{jalalikanafi_imprint_2024}, who found an uncertainty of $0.015$~eV. First, a direct comparison requires caution due to differing covariance estimations: our analysis employs 512 subboxes of $(350\,\mathrm{Mpc})^3$, whereas theirs utilizes 5000 simulations with a 1 $\mathrm{Gpc^3} h^{-3}$ box size, leading to substantially different volumes and sampling variance. Second, the non-Gaussianity of the posterior distribution is a plausible explanation for the further discrepancy with the Fisher forecasts. Finally, our analysis is generalized to a wider cosmological parameter space that also varies parameters like the dark energy equation of state, dark matter lifetime, and running of the spectral index.

We present the constraints derived using only the three intermediate persistence bins ($p = 1, 2, 3$) in Table~\ref{tab:bb_compare}. This conservative selection excludes the lowest and highest bins ($p = 0$ and $4$), which exhibited less stable performance in our emulator and, in the case of the lowest bin, greater susceptibility to shot noise. Even with this restricted dataset, we obtain a robust neutrino mass constraint with an uncertainty of $0.06\text{-}0.07~\text{eV}$. This result confirms that the primary cosmological signal does not reside in the lowest-persistence bin, reinforcing the practical applicability of our method to future observational data where such noise-sensitive regimes may be excluded.

\begin{figure*}
    \centering
    \includegraphics[width=0.95\linewidth]{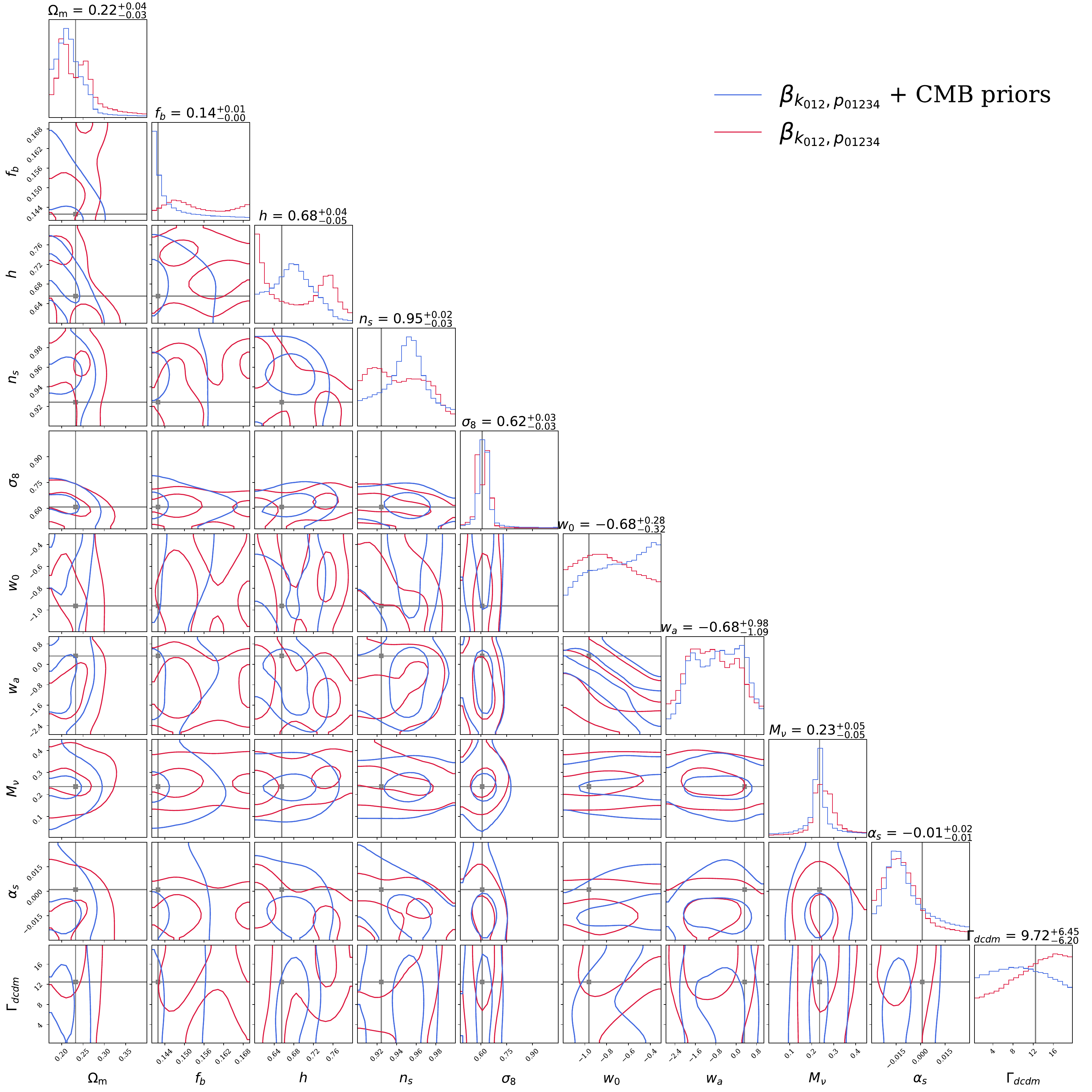}
    \caption{The contour plot of the constraints on all ten parameters in the emulator, using persistence strips with (blue contours) and without the CMB priors (red contours). The grey lines indicate the true value of each parameter. Labels in each column show the 1-$\sigma$ constraints with CMB priors.}
    \label{fig:Nau_Priors_results}
\end{figure*}

\subsubsection{Other parameters}

We present constraints on all the cosmological parameters from our topological statistics, comparing inferences with and without CMB priors in Fig.~\ref{fig:Nau_Priors_results}. 
Topology alone yields constraints on most parameters, though with varying levels of precision; however, the inclusion of CMB priors significantly improves their tightness. 
The most pronounced improvement is seen for the baryon fraction, $\fb$, which exhibits a substantial gain in both accuracy and precision. This is expected, as the CMB priors directly provide tight constraints on the physical baryon density $\rm \omega_b$ and the angular acoustic scale $\theta_{\star}$.

The CMB priors introduce a tight degeneracy between the dark energy equation of state parameters, $w_0$ and $w_a$, but the posteriors remain significantly biased from the true values. This bias likely stems from residual degeneracies with other poorly constrained parameters, such as $\ns$, which is itself biased. 
The spectral index $\ns$ is slightly improved by the CMB prior but remains outside the $1\sigma$ confidence level. The running of the spectral index $\alphas$ shows no significant improvement with the inclusion of CMB priors.
The decaying dark matter parameter $\Gammadcdm$ remains entirely unconstrained in both cases.

For the four key parameters, the CMB priors yield more nuanced effects:
The constraints on $\Omegam$, $\rm \sigma_8$, and $\SigmaMnu$ show marginal improvements.
The constraints on $\h$ improve significantly from a bimodal distribution to a clearly defined Gaussian peak.

\begin{table*}
    \centering
     \begingroup
\renewcommand{\arraystretch}{1.3} 
    
    \begin{tabular}{c |c c | c c c c}
    \hline\hline
    Probe &  CMB &  Field & $\Omegam$ & $\h$ & $\rm \sigma_8$ & $\SigmaMnu$/eV \\
&  priors & used & 0.233 & 0.655 & 0.607 & 0.236 \\
    \hline \hline
    $\beta_{k_{{012}},p_{123}}$ &
    \checkmark & F &
    $0.214_{-0.028}^{+0.042} $ & 
    $0.680_{-0.053}^{+0.047} $ & 
    $0.619_{-0.040}^{+0.045} $ & 
    $0.236_{-0.072}^{+0.062} $ 
    \\
    $\beta_{k_{{012}},p_{01234}}$ &
    \checkmark & F &
    $0.216_{-0.029}^{+0.042} $ & 
    $0.677_{-0.054}^{+0.048} $ & 
    $0.618_{-0.037}^{+0.035} $ & 
    $0.237_{-0.050}^{+0.051} $ 
    \\
     \hline
    $\beta_{k_{{012}},p_{123}}$ &
    \checkmark & D &
    $0.224_{-0.034}^{+0.030} $ & 
    $0.664_{-0.036}^{+0.056} $ & 
    $0.616_{-0.029}^{+0.051} $ & 
    $0.152_{-0.094}^{+0.176} $ 
    \\
    $\beta_{k_{{012}},p_{01234}}$ &
    \checkmark & D &
    $0.220_{-0.029}^{+0.031} $ & 
    $0.671_{-0.039}^{+0.048} $ & 
    $0.619_{-0.031}^{+0.037} $ & 
    $0.228_{-0.137}^{+0.132} $
    \\
    \hline
    $\beta_{k_{{012}}}$ &
    \checkmark & T &
    $0.212_{-0.024}^{+0.029} $ & 
    $0.681_{-0.036}^{+0.041} $ & 
    $0.621_{-0.030}^{+0.046} $ & 
    $0.254_{-0.094}^{+0.077} $ 
    \\
    1 pts &
    \checkmark & T &
    $0.244_{-0.055}^{+0.025} $ & 
    $0.640_{-0.032}^{+0.080} $ & 
    $0.580_{-0.046}^{+0.151} $ & 
    $0.215_{-0.133}^{+0.137} $ 
    \\
    1 pts + $\beta_{k_{{012}},p_{01234}}$ &
    \checkmark & T/D &
    $0.251_{-0.059}^{+0.016} $ & 
    $0.632_{-0.022}^{+0.085} $ & 
    $0.590_{-0.015}^{+0.063} $ & 
    $0.238_{-0.084}^{+0.073} $ 
    \\
    \hline\hline
    \end{tabular}
    \endgroup
    \caption{Similar to Table~\ref{tab:bb_compare}, but showing results from different fields and probes, including the topological analysis on the F, T, and D fields, as well as the 1-pts analysis of the T field and the joint analysis of 1-pts with topological analysis on field D.}
    \label{tab:bb_compare2}
\end{table*}

\subsection{Persistence strips over Betti curves}

\begin{figure}
    \centering
    \includegraphics[width=0.95\linewidth]{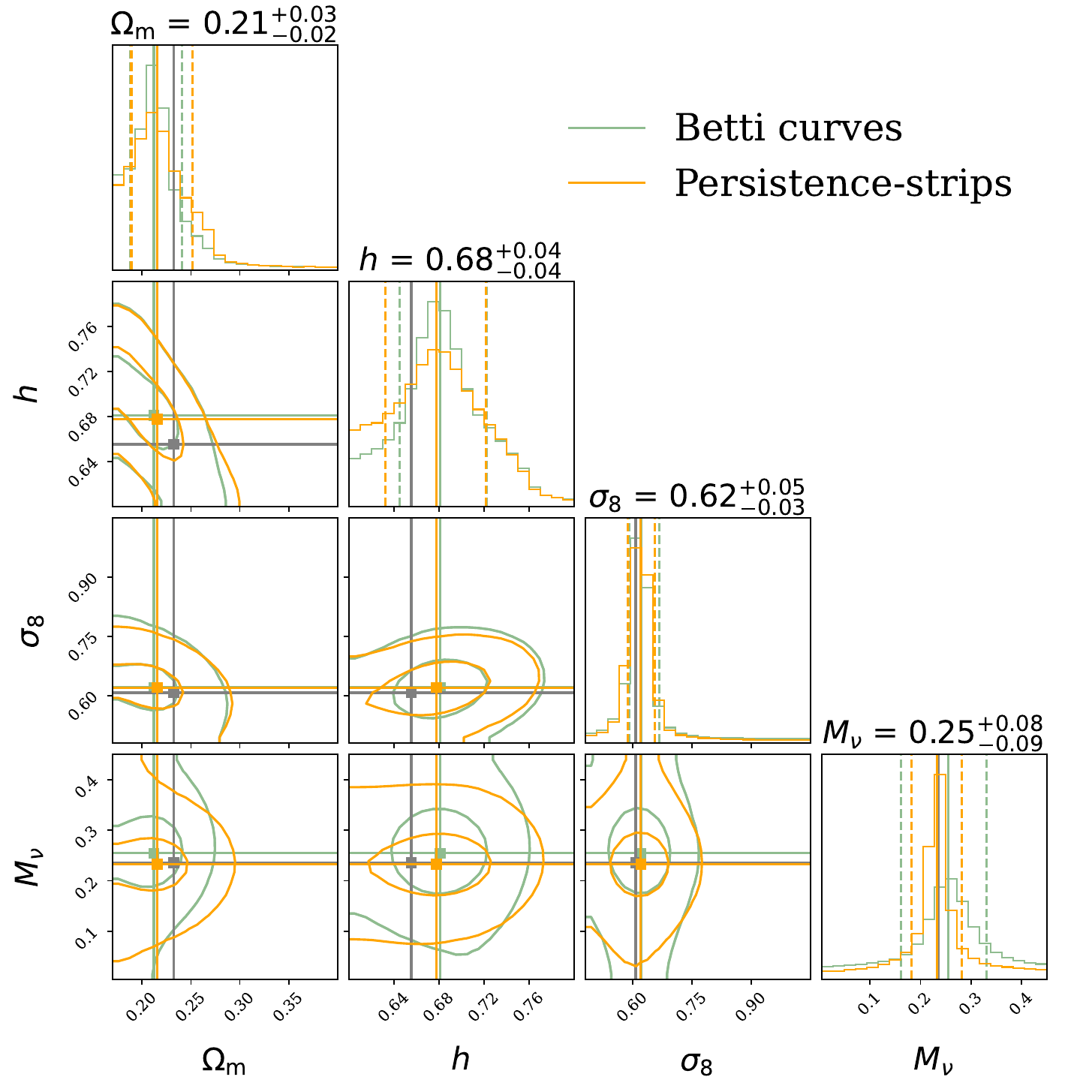}
    \caption{Similar to Fig.~\ref{fig:corner_bb_compare}, but comparing the cosmological inference from persistence strips (orange contours) with that from normal Betti curves (green contours), all with CMB priors. The grey lines indicate the true value of each parameter. Labels on top of each column indicate the 1$\sigma$ confidence level of the normal Betti curves.}
    \label{fig:Nau_Betti_compare}
\end{figure}

To maximize the information extracted from persistence diagrams and to facilitate the emulation of topological features, we have introduced the persistence strips framework. This approach transforms the complex topological data into a set of one-dimensional curves binned by persistence, which are more amenable to interpretation and emulation than the full diagrams.

To quantify the advantage of this persistence-based binning, we construct a separate emulator using the standard Betti curves from the same simulation suite. A comparison of the cosmological constraints from both methods is presented in Fig.~\ref{fig:Nau_Betti_compare}. While both the standard Betti curves and persistence strips successfully break the key degeneracy between $\SigmaMnu$ and $\rm \sigma_8$, the persistence strips achieve about two times tighter constraints on $\SigmaMnu$ (see Table~\ref{tab:bb_compare2}). 
The marginally weaker constraints on $\Omega_\mathrm{m}$ and $\sigma_8$ from the persistence strips as shown in Fig.~\ref{fig:Nau_Betti_compare} occurred in only a single instance among our ten test samples. Generally, 
the two probes show similar performance.
In contrast, the improvement in the constraint for the neutrino mass $\SigmaMnu$ was systematic and robust across the entire test set.

This advantage is expected to be even more pronounced in real observational data. In such cases, density fluctuations and observational noise can significantly bias low-persistence structures. The inherent ability of persistence strips to separate the robust, high-persistence signal from the noisy, low-persistence features will therefore be critical for robust cosmological inference.

\subsection{Inference from dark matter-only field}
\begin{figure}
    \centering
    \includegraphics[width=0.95\linewidth]{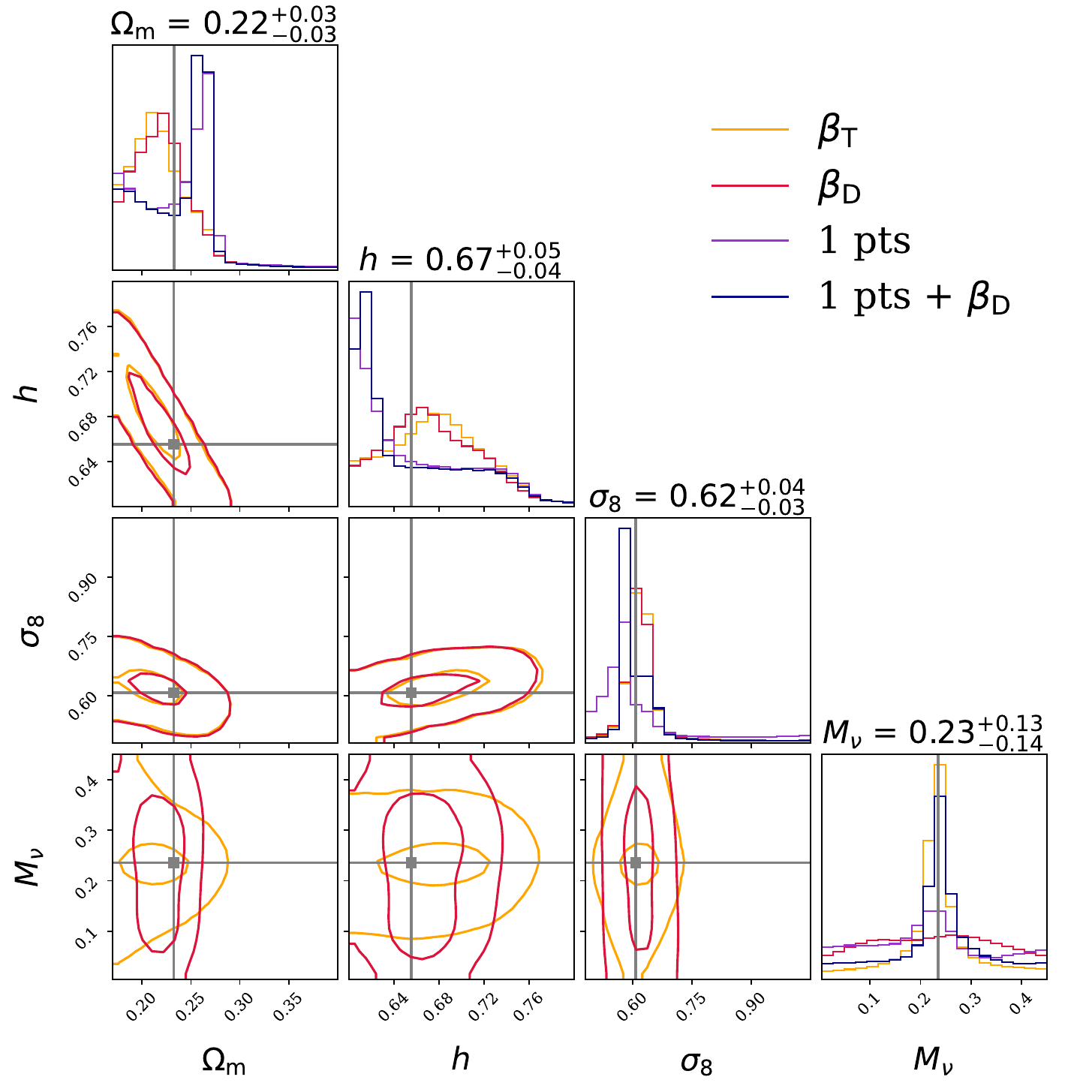}
    \caption{Similar to Fig.~\ref{fig:corner_bb_compare}, but comparing the results from T and D fields. For clarity, we only show the confidence contours from the T and D fields' topology (with CMB priors, in orange and red, respectively) in 2-D subpanels. 
    We also include the results from 1-pts and the joint analysis of 1-pts with field D in the 1-D subpanels (purple and navy respectively), with CMB priors. The grey lines indicate the true value of each parameter. Labels above each column indicate the median and 68\% credible intervals for $\beta_\mathrm{D}$.}
    \label{fig:DM}
\end{figure}

Notably, when applying the emulator to the dark matter-only (D) field, we observe significantly weaker constraints on $\sum m_\nu$ across all persistence strips compared to the distinct signal identified in the total matter (T) field, as shown in Fig.~\ref{fig:DM}. The constraint obtained from the D field is $0.13$~eV, which aligns more closely with the Fisher forecast of $0.12$~eV reported in \citet{jalalikanafi_imprint_2024}. 
In contrast, constraints on other parameters, such as $\sigma_8$ and $\Omega_\mathrm{m}$, do not show significant differences between the D and T fields.

We hypothesize that this discrepancy arises from a stronger degeneracy between neutrino mass and other cosmological parameters in the D field, which hinders the ability of topological statistics to cleanly isolate the neutrino signature. This interpretation is consistent with the findings of \citet{jalalikanafi_imprint_2024}. We further explore the physical origin of this degeneracy in Section~\ref{sec:discuss}.

\section{Physical Origins of Neutrino Mass Sensitivity}\label{sec:discuss}
A key objective of this work is to understand the physical mechanism that breaks this cosmological degeneracy with the `all-matter' field T. In this section, we discuss and check the possible physical reasons for the sensitivity of the T-field's persistence strips to neutrino mass. We conclude that the overall neutrino mass fraction and the distribution of the cold dark matter are equally important, while the distribution of the neutrinos is less important.

\subsection{Neutrino fraction or field?}
As can be seen in Fig.~\ref{fig:mul-density}, the contribution of the distribution of neutrinos is subdominant compared to that of the suppression of the D field.
To isolate the effect of the distribution of neutrinos, we construct a `Fake all-matter' field. This field is identical to the `True all-matter' field except that we replace the true neutrino field with its mean value, i.e. $\delta_\nu=0$.
The field F only depends on neutrino masses in two ways: through the global average neutrino density, as a fraction of the total matter density, and through the indirect impact on the distribution of dark matter.
This setup allows us to compare four distinct density fields to precisely determine the source of the constraining power. The cold dark matter (D) and neutrino (N) fields are simply
\begin{align}
\rm 1+\delta_D &= 1+ \delta_{\rm cdm},\\
\rm 1+\delta_N &= 1+ \delta_{\nu}.
\end{align}

\noindent
The true total matter field (T) is their weighted average:
\begin{align}
\begin{split}
\rm 1+\delta_T& = 1+\frac{\Omega_{\rm cdm}}{\Omegam}\delta_{\rm cdm} + \frac{\Omega_{\nu}}{\Omegam} \delta_{\nu} \\ 
        & = (1+\delta_{\rm cdm}) - \rm \frac{\Omega_{\nu}}{\Omegam} (\delta_{cdm} - \delta_{\nu}),
\end{split}
\end{align}

\noindent
whereas the fake total matter field (F) is defined as
\begin{align}
\begin{split}
\rm 1+\delta_F &=  1+\frac{\Omega_{\rm cdm}}{\Omegam} \delta_{\rm cdm} \\
        &= (1+\delta_{\rm cdm}) - \rm \frac{\Omega_{\nu}}{\Omegam} \delta_{cdm}.
\end{split}\label{eq:fields}
\end{align}

\begin{figure}
    \centering
    \includegraphics[width=0.95\linewidth]{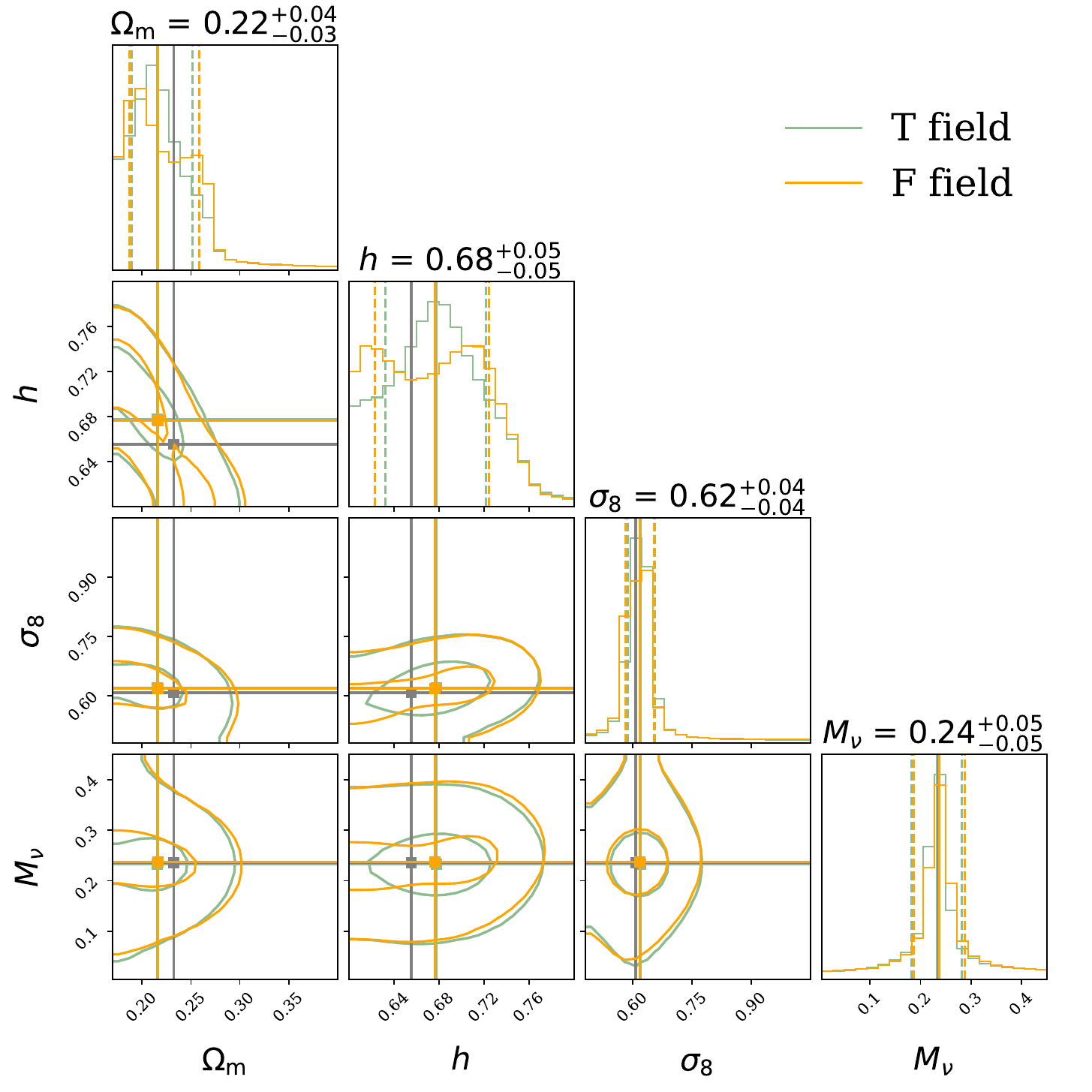}
    \caption{Similar to Fig.~\ref{fig:corner_bb_compare}, but comparing the cosmological inference from field T (green contours) with that of field F (orange contours), all with CMB priors. The grey lines indicate the true value of each parameter.}
    \label{fig:TF}
\end{figure}

We construct an emulator based on field F following the same procedure as in Section.~\ref{sec:emu}. The parameter dependence and cosmological inference on the four main parameters $\Omegam$, $\h$, $\rm \sigma_8$ and $\SigmaMnu$ are very close to those for Field T (See Fig.~\ref{fig:TF} and Table~\ref{tab:bb_compare2}). We show the cosmological constraints from field T and field F in Fig.~\ref{fig:TF}. The figure demonstrates that the constraining power remains similar, indicating that the information does not derive from the distribution of neutrinos but rather from the neutrino fraction in the total matter density and the topology of the cold dark matter.
In other words, these results show that the neutrino fraction rather than the neutrino field contributes to revealing the neutrino mass and breaking parameter degeneracies. The fluctuations in the neutrino density field itself provide only a minor contribution to the topological imprint of neutrino mass.

\subsection{Mathematical explanation}
This behaviour can be explained mathematically by considering the definition of the fields in Eq.~\ref{eq:fields}. Under a simplified assumption that the cold dark matter density contrast, $\delta_{\rm cdm}$, remains approximately constant for different neutrino masses, the variation in Field F is primarily driven by the neutrino fraction, $\Omega_{\nu}/\Omega_\mathrm{m}$. 
A higher neutrino mass causes matter density fluctuations to be suppressed, driving structures closer to the mean density ($\delta=0$); overdense regions become less dense, and underdense regions become less empty.
This differential shift contributes significantly to the observed `tightening' pattern in the persistence distributions (for curves expanding across the point $\vartheta_{\rm death} = 1$, in this case, $\beta_0$ and $\beta_1$) and the shift in peak density (for curves with $\vartheta_{\rm death} < 1$).

This nonlinear transformation of the density field also accounts for the observed suppression of high-persistence voids and enhancement of low-persistence ones. 
For example, a higher neutrino mass causes a topological feature crossing the $\delta=0$ threshold to experience a decrease in $\vartheta_{\rm birth}>0$ and an increase in $\vartheta_{\rm death}<0$, pushing it toward lower persistence rather than simply rescaling its density thresholds.
Thus, the enhanced sensitivity of void topology ($\beta_2$) to neutrino mass, compared to halos ($\beta_0$) and filaments ($\beta_1$), arises largely from the relatively higher fraction of neutrinos.


In a more realistic scenario, $\delta_{\rm cdm}$ is also modulated by the neutrino mass, due to the suppression of the growth of CDM perturbations. Heavier neutrinos lead to a more clustered neutrino field and, in contrast, a less clustered cold dark matter field. This is the origin of the degeneracy between $\SigmaMnu$ and $\rm \sigma_8$ within the field D. The question remains how much information is derived from these two effects: the change in the neutrino fraction and the suppression of the D field.

\subsection{Two effects with equal importance}

Based on these results, we considered filtrations as a function of the volume fraction instead of the overdensity, following the rescaling method proposed by \citet{Weinberg_rescale_1988}, thereby removing information from the one-point distribution function (1pts hereafter) of the density field, $P(\delta)$. Under this transformation, the Betti curves for the T field become nearly indistinguishable from those of the D field. This indicates that the enhanced constraining power of the total matter field originates not from a fundamentally different topology, but from the way that neutrinos modify the underlying 1pts of the T field.
This is consistent with the results of \citet{Liu_onepoint_2019}, showing that 1pts are capable of constraining the neutrino mass.
The case is more complex for persistence strips. Their dependence on structure lifetimes (persistence) cannot be straightforwardly rescaled by this method, suggesting that they encode information beyond the 1pts. This likely explains why persistence strips achieve approximately twice the constraining power on $\sum m_\nu$ compared to conventional summary statistics.

To isolate the constraining power arising purely from changes to $P(\delta)$, we build an emulator of the one-point probability distribution function following the same methodology and present the results in Fig.~\ref{fig:DM}. Both the 1pts and D-field topology individually yield weak constraints on neutrino mass. However, only by combining these two probes can we closely match the constraining power achieved with the T‑field topology alone.
This confirms that persistent homology captures information beyond simple field rescaling and these two effects are equally important. 

\section{Observational considerations}\label{sec:obs}

To assess the robustness of our findings under more realistic observational conditions, we incorporate the effect of redshift-space distortions (RSD) and test against different smoothing scales. 

\subsection{Smoothing scale effect}\label{sec:smoothing}
\begin{figure}
    \centering
    \includegraphics[width=0.95\linewidth]{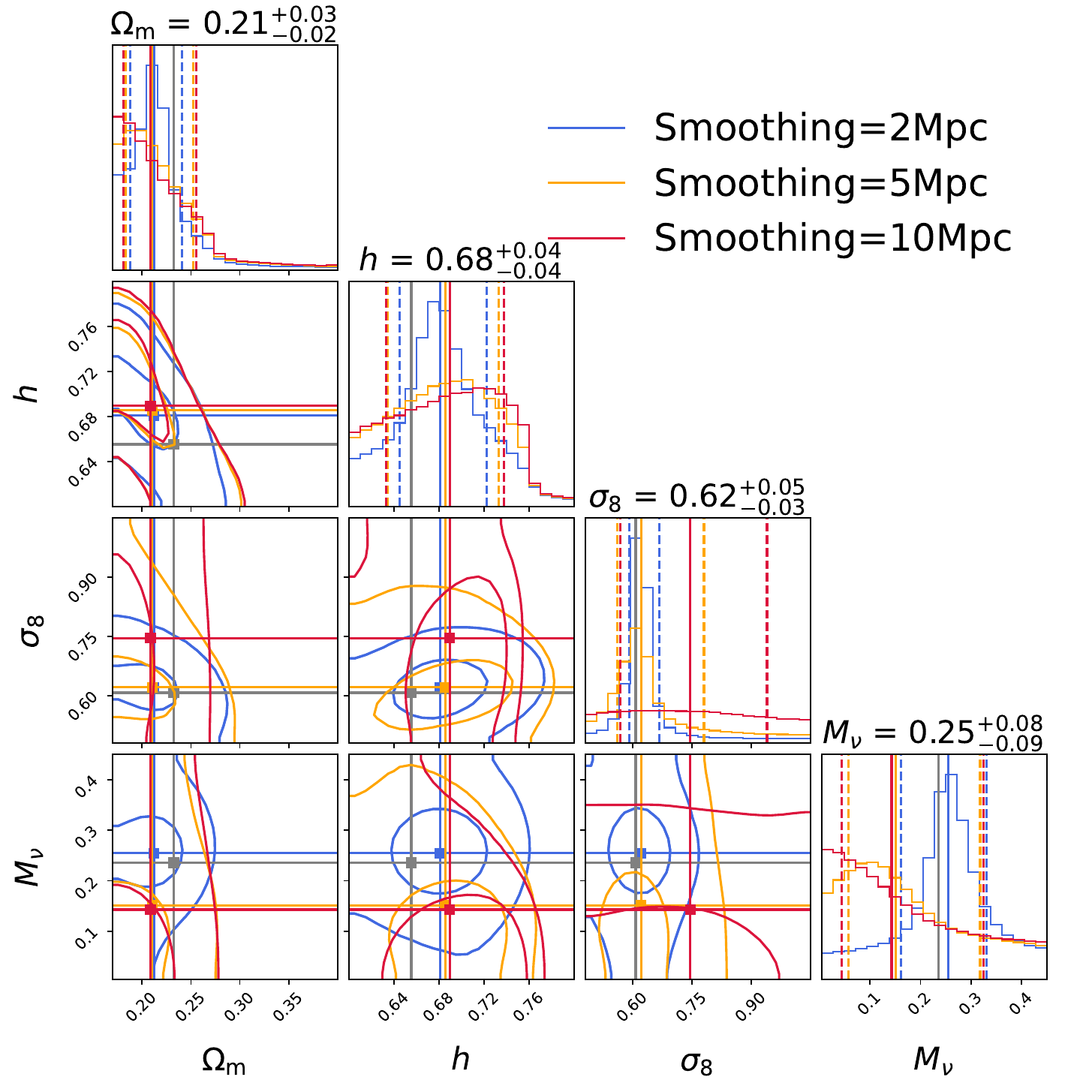}
    \caption{Similar to Fig.~\ref{fig:corner_bb_compare}, but comparing the cosmological inference from Betti curves obtained from fields with different smoothing scales, $2\mpc$ (blue), $5\mpc$ (orange), and $10\mpc$ (red), all with CMB priors. The grey lines indicate the true value of each parameter.}
    \label{fig:Nau_smoothscale_noslice}
\end{figure}
\citet{jalalikanafi_imprint_2024} demonstrated that increasing the smoothing scale degrades constraints on neutrino mass, reporting a 68\% confidence interval for $\SigmaMnu$ that widens from $0.0152$~eV with a $5\mpch$ kernel to $0.0720$~eV with a $10\mpch$ kernel.
In this work, the $\delta f$ method combined with the higher simulation resolution enables us to probe significantly smaller scales, down to $2\mpc$.
To systematically investigate the smoothing scale dependence of our constraints, we construct two additional emulators for comparison with our $2\mpc$ baseline, employing smoothing scales of $5\mpc$ and $10\mpc$ while keeping all other procedures identical. We note that the meaningful persistence range changes substantially with the smoothing scale, making it difficult to quantify this difference and introducing challenges for a fair comparison of the persistence strips across different configurations. Although we present the complete results for both Betti curves and persistence strips in Table~\ref{tab:bb_coh}, the following comparative analysis focuses primarily on the Betti curves to enable a more straightforward interpretation.

The resulting constraints from all three smoothing scales are compared in Fig.~\ref{fig:Nau_smoothscale_noslice}. We find that different smoothing scales generally yield consistent cosmological predictions for $\Omega_\mathrm{m}$, $h$, and $\sigma_8$, but with degraded precision. Specifically, the constraining power on $\sigma_8$ almost disappears when increasing to $10\mpc$ smoothing. The constraints on $\SigmaMnu$ are largely biased to smaller values and lose constraining power with larger smoothing scales, indicating a large smoothing may erase the neutrino imprints. We have verified that the Bayesian inference is well-converged and consistent across several testing samples, ruling out sampling issues as the cause.

This degradation in cosmological constraining power with a larger smoothing scale can be attributed to its effect on cosmic voids. As the smoothing scale increases, it erodes the topological signature of voids. Given that our primary constraining power originates from high-persistence topological features, this erosion of void signatures provides a natural explanation for the significantly weaker cosmological constraints we obtain.

\subsection{Redshift space distortion effect}
\begin{figure}
    \centering
    \includegraphics[width=0.95\linewidth]{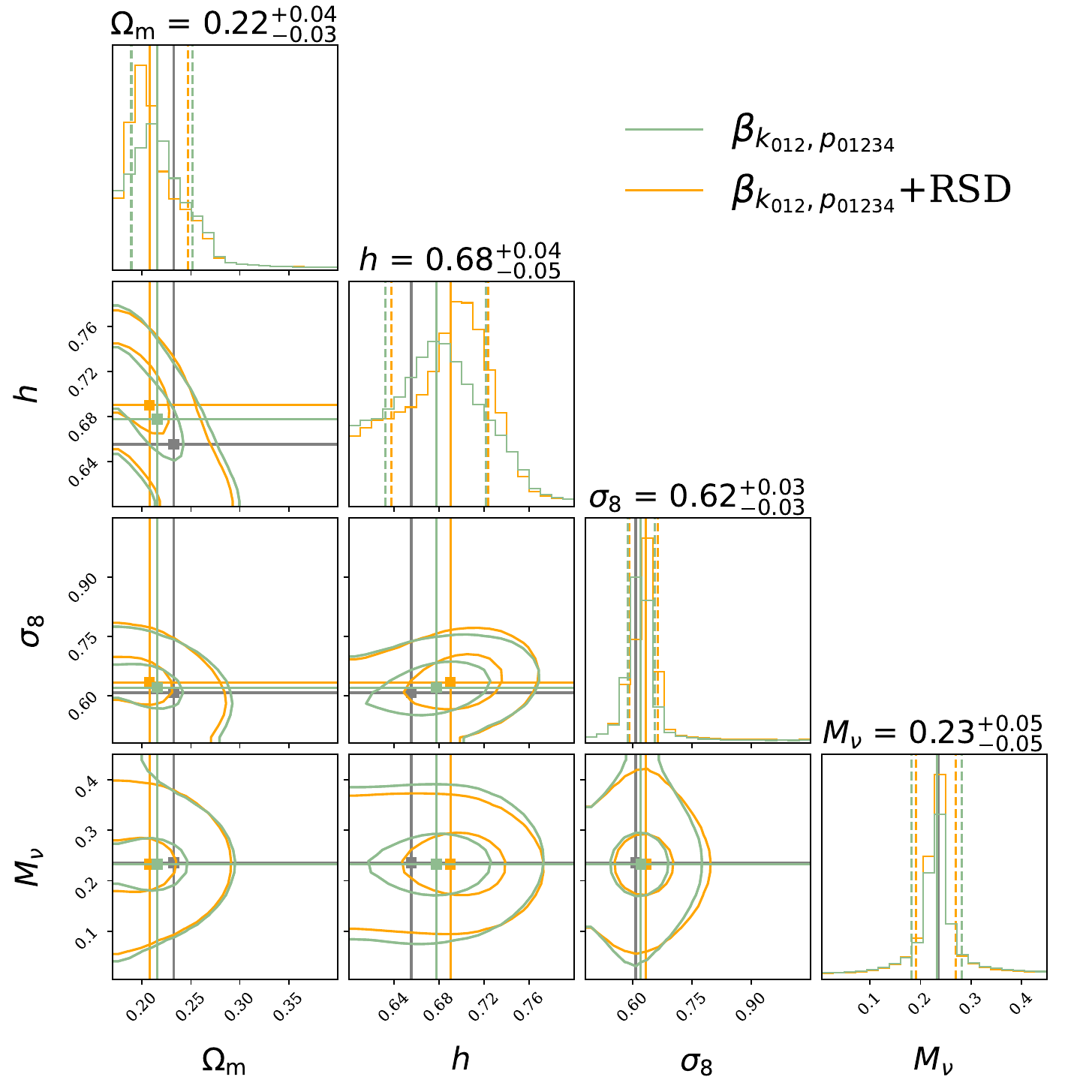}
    \caption{Similar to Fig.~\ref{fig:corner_bb_compare}, but comparing the cosmological constraints in real and redshift space for the joint analysis with persistence strips combined with CMB priors. Green and orange contours show the 68\% and 95\% confidence regions for real space and redshift space analyses respectively. The true parameter values are indicated by grey lines.}
    \label{fig:Nau_RSD_results}
\end{figure}

To assess the robustness of our method under more realistic observational conditions, we incorporate redshift-space distortions (RSD) into the dark matter particles in the simulation and repeat the cosmological inference procedure. 
The comoving position of each dark matter particle in redshift space, $\mathbf{s}$, is calculated from its real-space position, $\mathbf{x}$, and peculiar velocity, $\mathbf{v}$, using the standard transformation
\begin{align}
\mathbf{s}=\mathbf{x}+\frac{\mathbf{v} \cdot \hat{\mathbf{n}}}{a H(a)} \hat{\mathbf{n}},
\end{align}
where $\hat{\mathbf{n}}$ is the unit vector along the line-of-sight direction, $a$ is the scale factor, and $H(a)$ is the Hubble parameter at that epoch. 
We implement RSD using the plane-parallel approximation, which is well-justified for our simulation geometry with the observer effectively at infinity. The line-of-sight direction $\hat{\mathbf{n}}$ is aligned with one principal axis ($z$-axis) of the simulation box. For each dark matter particle, we use the line-of-sight component of its peculiar velocity and apply the corresponding displacement, scaled by the Hubble expansion factor $aH(a)$, accounting for periodic boundary conditions.
The resulting redshift-space dark matter density field is then constructed from the displaced particle positions using the same CIC mass assignment scheme as applied to the real-space density field.

The resulting constraints, presented in Fig.~\ref{fig:Nau_RSD_results}, demonstrate that our topological approach is robust to RSD effects: the constraints on the four most prominent parameters generally remain unchanged. 
The introduction of RSD shifts the constraints along existing degeneracy directions toward lower $\Omega_\mathrm{m}$ and higher $\h$, while offering essentially the same constraint on these parameters. 
Crucially, the constraint on the neutrino mass $\SigmaMnu$ remains robust, and the Betti curves continue to effectively break the degeneracy between $\SigmaMnu$ and other cosmological parameters.
These results confirm that topological statistics remain a powerful tool for constraining neutrino mass even in redshift space, providing strong support for the application of this method to real observations.

\subsection{Other considerations}

When addressing realistic observational effects, it is essential to recognize that galaxy surveys probe the discrete distribution of galaxies, not the continuous dark matter field. Persistent homology of discrete tracers are based on simplicial complexes \citep[e.g.][]{Bermejo2024} and may introduce several interrelated effects that are critical for interpreting galaxy survey data. 

We should note a series of tracer-dependent effects. At the most basic level, we observe that more massive halos are rarer, so only sufficiently large topological features (e.g., giant voids) are well delineated, while smaller-scale filaments and walls become poorly resolved.  
Second is the clustering bias effect, whereby massive halos cluster more strongly and preferentially occupy high‑density regions, leading to voids that are systematically larger and more regular than those in the full matter distribution. 
Most subtly, \citet{Bermejo2024} identify a distinct topological bias effect, wherein even after controlling for mass function and clustering, different halo populations trace systematically different topological features. In particular, more massive halos consistently probe larger‑scale structures.

Another viable technical approach involves conducting topological analyses directly on weak lensing convergence maps. However, this method faces a primary challenge of information loss, since the convergence map essentially projects the line-of-sight structure onto a two-dimensional plane, which discards spatial information along the redshift dimension (although the limited three-dimensional information provided by tomography offers a partial resolution). Nevertheless, works like \citet{Prat25_DES} have demonstrated the feasibility and potential of this approach. Its application to constraining neutrino mass remains an open question, which we will explore further in future work.

\section{Summary} \label{sec:summary}

This work establishes the cosmological constraining power of TDA, specifically through persistent homology, as a superior alternative to traditional two-point statistics for probing neutrino mass and breaking parameter degeneracies. 
Utilizing the FLAMINGO simulation suite, we developed a robust emulator spanning a 10-dimensional $w_0 w_a\text{CDM} +\nu$ cosmological parameter space.
Trained on only 50 simulations, it accurately predicts topological summaries (Betti curves and persistence diagrams) and enables Bayesian inference.

Our key findings show that Betti curves subdivided by persistence (the `persistence strips') exhibit enhanced sensitivity to the sum of neutrino mass, $\SigmaMnu$. Persistence strips strike an optimal balance between information content and emulation feasibility, delivering roughly twice the constraining power of unbinned Betti curves. The strongest constraints come from two‑dimensional topological structures (voids), which exhibit a characteristic response to neutrino diffusion. 
Using the emulator trained on only 50 simulations, each with a volume of $(350\, \mpc)^3$, we obtain a 68\% uncertainty of $0.05$~eV, demonstrating the superior sensitivity of persistent homology. 

The total matter field yields substantially tighter constraints ($0.05$~eV) than the dark‑matter‑only field ($0.13$~eV), underscoring the importance of probes sensitive to the total matter distribution (e.g., weak lensing). We performed a detailed investigation of the source of this enhancement. By analyzing a ``fake'' total matter field, in which the neutrino density perturbations have been replaced with their mean value, we established that variations in the global neutrino mass fraction, rather than the direct imprint of the neutrino perturbations, dominate the signal. Furthermore, by constructing emulators of multiple probes, we demonstrated that the sensitivity of the total matter topology can be decomposed into two effects of comparable importance: the indirect effect of neutrino mass on the growth of cold dark matter perturbations and the imprint of the neutrino mass fraction on the 1-point distribution of the total matter field.

The method proves robust to redshift‑space distortions, converges with 50 simulations, and effectively breaks degeneracies among $\SigmaMnu$ and $\sigma_8$, paving the way for its application to upcoming observational data from DESI and other wide-field surveys.
Looking forward, we leave the inspection of hydrodynamical effects and the application of our formalism to realistic galaxy mocks and weak lensing maps, incorporating the characteristics of surveys like Euclid and DESI, for future work.

\section*{Acknowledgements}
The authors acknowledge useful discussions with Carlton Baugh and Hailiang Du on the construction of the GP emulator.
The authors also acknowledge support from the China Scholarship Council, the Science and Technology Facilities Council (grant numbers ST/X001075/1 and ST/Y002733/1), the National Key R\&D Program of China (2023YFA1607800, 2023YFA1607804), and “the Fundamental Research Funds for the Central Universities”, 111 project No. B20019. RvdW acknowledges funding from EU Horizon Europe (EXCOSM, grant nr. 101159513).
This work used the DiRAC@Durham facility managed by the Institute for Computational Cosmology on behalf of the STFC DiRAC HPC Facility (www.dirac.ac.uk). The equipment was funded by BEIS capital funding via STFC capital grants ST/K00042X/1, ST/P002293/1, ST/R002371/1 and ST/S002502/1, Durham University and STFC operations grant ST/R000832/1. DiRAC is part of the National e-Infrastructure.
Part of the computations in this paper were also run on the the Gravity Supercomputer at the Department of Astronomy, Shanghai Jiao Tong University. 

\section*{Data Availability}
The data underlying this article will be shared on reasonable request to the corresponding author.

\bibliographystyle{mnras}
\bibliography{ref}

\appendix
\setcounter{figure}{0}
\setcounter{table}{0}

\section{The emulator convergence test}\label{app:emulator_number}

\begin{figure}
  \centering
      \includegraphics[width=0.95\linewidth]{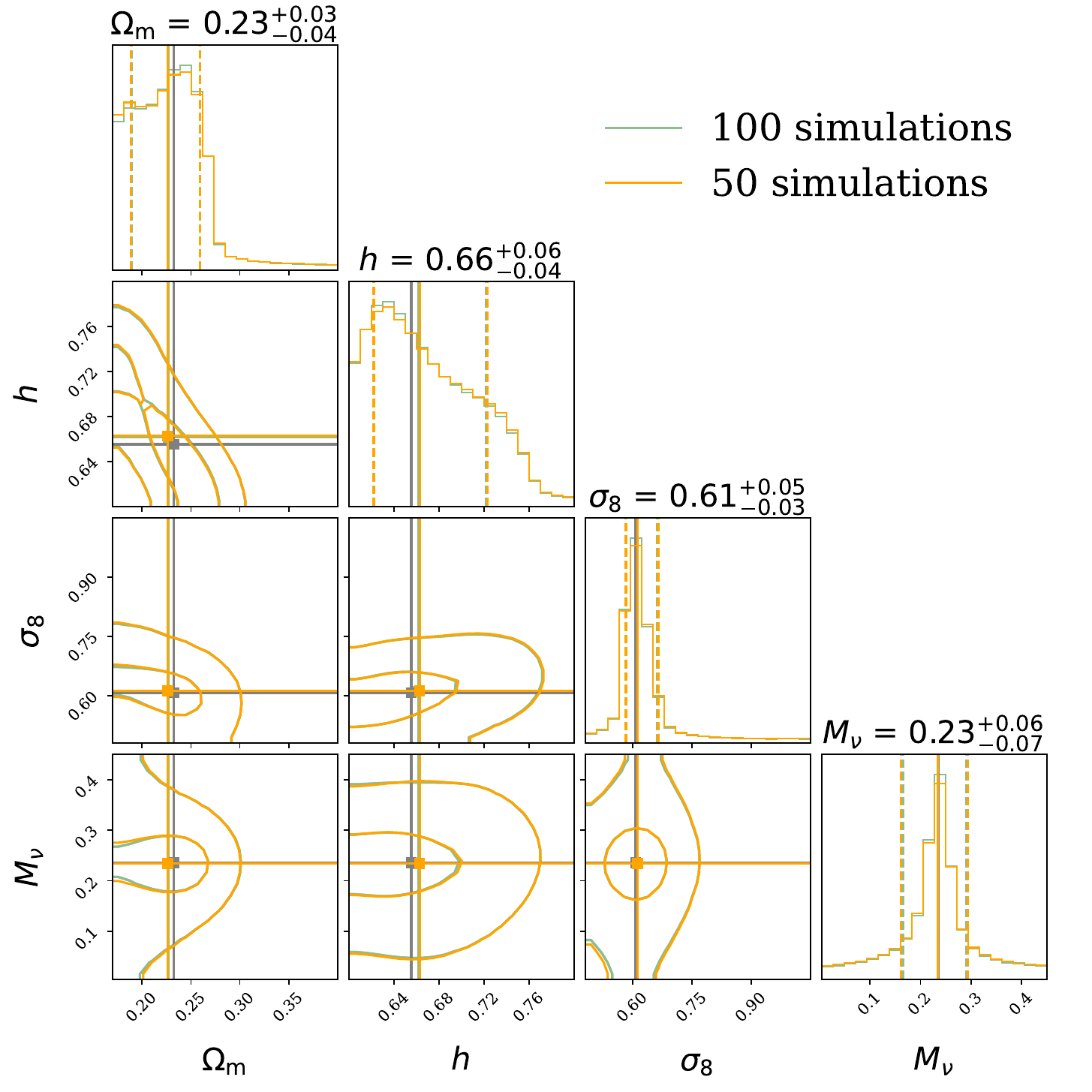}
      \caption{Similar to Fig.~\ref{fig:corner_bb_compare}, but comparing the results for different training sets. The constraints are derived from ensembles of 50 and 100 simulations, all with CMB priors. The grey lines indicate the true value of each parameter.}\label{fig:50_100}
\end{figure}

We assess the convergence of our emulator by comparing constraints derived from ensembles of 50 and 100 simulations. As shown in Fig.~\ref{fig:50_100}, the resulting cosmological constraints from the two ensembles are nearly identical. This indicates that the emulator built on 50 simulations has sufficiently converged, with a negligible gain in precision from doubling the size of the training set.

\section{Emulator with different cosmological parameters}\label{app:emulator}

\begin{figure}
      \centering
      \includegraphics[width=0.95\linewidth]{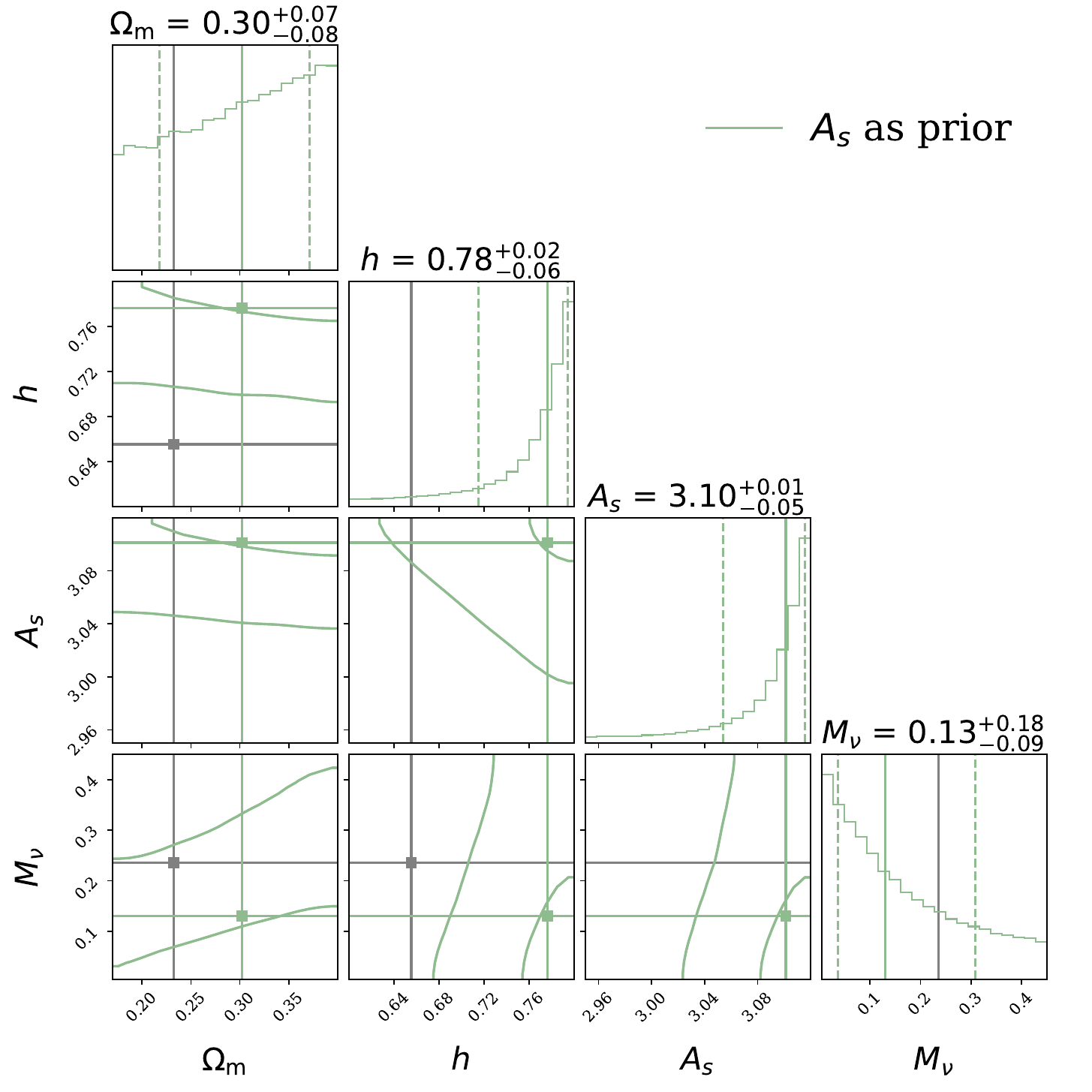}
    \caption{Similar to Fig.~\ref{fig:corner_bb_compare}, but with $A_s$ as prior. The grey lines indicate the true value of each parameter.}
    \label{fig:Nau_conv_As}
\end{figure}

\begin{table*}
    \centering
     \begingroup
\renewcommand{\arraystretch}{1.3} 

     \begin{tabular}{c |c c c c | c c c c}
    \hline\hline
        Probe & Smoothing & CMB &  Field & parameters & $\Omegam$ & $\h$ & $\rm \sigma_8$ & $\SigmaMnu$ \\
& & scale (Mpc) & priors & used  & 0.233 & 0.655 & 0.607 & 0.236
        \\
        \hline \hline
    $\beta_{k_{{012}},p_{123}}$ &
    5 & \checkmark & T &$\sigma_8$ &
    $0.251_{-0.041}^{+0.014} $ & 
    $0.630_{-0.016}^{+0.061} $ & 
    $0.588_{-0.005}^{+0.009} $ & 
    $0.248_{-0.136}^{+0.116} $
    \\
    $\beta_{k_{{012}},p_{123}}$ &
    10 & \checkmark & T &$\sigma_8$ &
    $0.212_{-0.030}^{+0.015} $ & 
    $0.685_{-0.022}^{+0.052} $ & 
    $0.639_{-0.024}^{+0.033} $ & 
    $0.397_{-0.272}^{+0.028} $
    \\
    \hline
    $\beta_{k_{{012}}}$ &
    2 & \checkmark & T &$\sigma_8$ &
    $0.212_{-0.024}^{+0.029} $ & 
    $0.681_{-0.036}^{+0.041} $ & 
    $0.621_{-0.030}^{+0.046} $ & 
    $0.254_{-0.094}^{+0.077} $ 
    \\
    $\beta_{k_{{012}}}$ &
    5 & \checkmark & T &$\sigma_8$ &
    $0.210_{-0.027}^{+0.042} $ & 
    $0.686_{-0.052}^{+0.047} $ & 
    $0.621_{-0.060}^{+0.159} $ & 
    $0.151_{-0.094}^{+0.167} $
    \\
    $\beta_{k_{{012}}}$ &
    10 & \checkmark & T &$\sigma_8$ &
    $0.209_{-0.027}^{+0.047} $ & 
    $0.689_{-0.057}^{+0.048} $ & 
    $0.746_{-0.177}^{+0.194} $ & 
    $0.143_{-0.099}^{+0.182} $ 
    \\
    \hline
    $\beta_{k_{{012}},p_{01234}}$ &
    2 &   & T & $\sigma_8$ &
    $0.227_{-0.037}^{+0.046} $ & 
    $0.682_{-0.068}^{+0.080} $ & 
    $0.616_{-0.037}^{+0.042} $ & 
    $0.252_{-0.063}^{+0.072} $  
    \\
    $\beta_{k_{{012}},p_{01234}}$ &
    2 &  & T & $A_s$ &
    $0.302_{-0.085}^{+0.069} $ & 
    $0.776_{-0.061}^{+0.018} $ & 
    $3.101_{-0.047}^{+0.014} $ & 
    $0.131_{-0.094}^{+0.178} $  
    \\
    \hline\hline
    \end{tabular}
    \endgroup
    \caption{Similar to Table~\ref{tab:bb_compare}, but with a different smoothing scale or density fluctuation parameter.}
    \label{tab:bb_coh}
\end{table*}
An intriguing finding emerges from the comparison of fluctuation amplitude parameters. While the amplitude of the primordial power spectrum, $A_s$, and the late-time fluctuation amplitude, $\sigma_8$, are traditionally degenerate and equally effective indicators for the power spectrum, this equivalence does not hold for topological statistics. We find that Betti curves exhibit a strong, well-defined response to $\sigma_8$, but only a weak and poorly constrained response to $A_s$ (see Fig.~\ref{fig:Nau_conv_As} and Table~\ref{tab:bb_coh}).

Consequently, this work utilizes $\sigma_8$ as the primary parameter for characterizing matter fluctuations. Since our simulation hypercube was designed with a flat prior on $\ln A_s$, the derived values of $\sigma_8$ follow a distribution that is approximately Gaussian. 
To mitigate potential biases from sparsely sampled regions of the parameter space, we restrict our testing samples to the range $\sigma_8 \in [0.6, 0.9]$.

\section{Betti curve dependence on other cosmological parameters}
\label{app:cosmo}
\begin{figure*}
    \centering
    \includegraphics[width=0.8\linewidth]{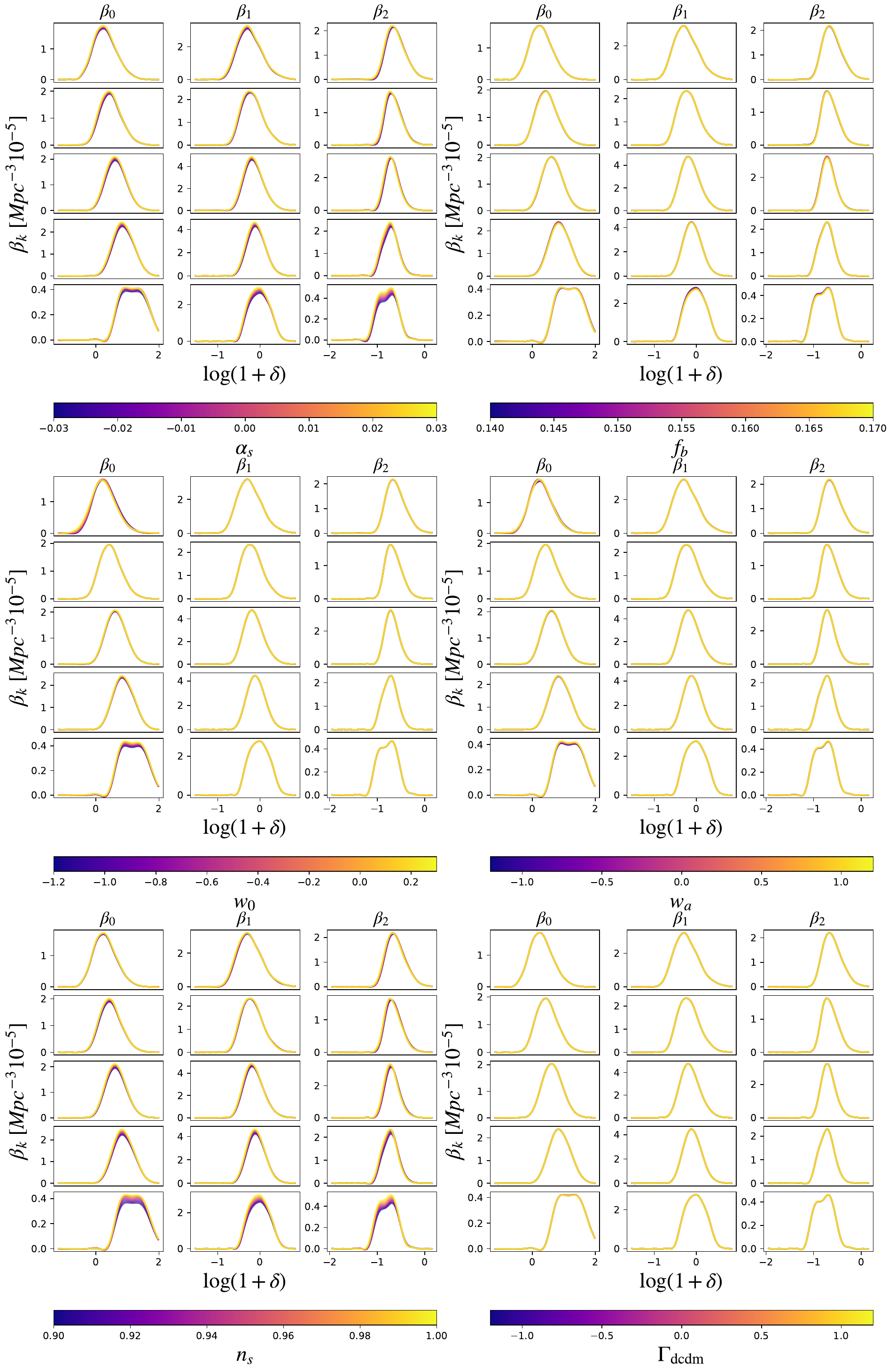}
    \caption{Similar to Fig.~\ref{fig:dependence}, but showing the dependence on 6 less prominent parameters.}
    \label{fig:lessimportantdependence}
\end{figure*}

The cosmological dependence of six less prominent parameters: the baryon fraction $f_\mathrm{b}$, the scalar spectral index $n_s$, its running $\alpha_s$, the decaying dark matter rate $\Gamma_\mathrm{dcdm}$, and the dark energy equation of state parameters ($w_0$, $w_a$) are shown in Fig.~\ref{fig:lessimportantdependence}. A measurable dependence is observed primarily for $\alpha_s$ and $n_s$ (most notably in high-persistence regimes), followed by a weaker dependence on $w_0$ and $w_a$. This indicates that the topology of the cosmic web is subtly influenced by both the initial conditions of the power spectrum and the dynamics of dark energy. Conversely, the topological statistics exhibit minimal sensitivity to the baryon fraction $f_\mathrm{b}$ and cannot meaningfully constrain the decaying dark matter parameter $\Gamma_\mathrm{dcdm}$.

\bsp	
\label{lastpage}
\end{document}